\documentclass[usenatbib]{mnras}
\usepackage{epsfig}
\usepackage{times}
\usepackage{amsmath,amssymb}
\usepackage{xcolor}
\usepackage{tikz}
\graphicspath{{figures/}}
\usepackage{cprotect}

\newcommand{\Msolar}{\mbox{\,$\rm M_{\odot}$}}        
\newcommand{\Lsolar}{\mbox{\,$\rm L_{\odot}$}}        

  \newcommand{\Teff}{\mbox{\,\em T$_{\rm eff}$}}         
 \newcommand{\teff}{\mbox{\,$T_{\rm eff}$}}      

  \newcommand{\kmsec}{\,\mbox{$\mbox{km}\,\mbox{s}^{-1}$}}    
  \def\simge{\mathrel{\raise1.16pt\hbox{$>$}\kern-7.0pt
    \lower3.06pt\hbox{{$\scriptstyle \sim$}}}}           
  \def\simle{\mathrel{\raise1.16pt\hbox{$<$}\kern-7.0pt
    \lower3.06pt\hbox{{$\scriptstyle \sim$}}}}           

\newcommand{\appropto}{\mathrel{\vcenter{
  \offinterlineskip\halign{\hfil$##$\cr
    \propto\cr\noalign{\kern2pt}\sim\cr\noalign{\kern-2pt}}}}}

\usepackage{tabu}   
\newcolumntype{L}[1]{>{\raggedright\let\newline\\\arraybackslash\hspace{0pt}}m{#1}}
\usepackage{afterpage}
\usepackage{caption} 
\usepackage{graphicx} 
\DeclareGraphicsExtensions{.pdf,.png,.jpg,.eps,.ps}
\usepackage{breqn}   
\usepackage{soul}  

\title[Kinematics and luminosities of EHe stars ]{The distribution, kinematics and luminosities of extreme  helium stars as probes of their origin and evolution}

\author[A.~{Philip Monai}, P.~Martin, C.~S.~Jeffery]{{A.~{Philip Monai}$^{1,2}$\thanks{email: asish.philip.monai@armagh.ac.uk}}, {P.~Martin$^{1,3}$, {C.~S.~Jeffery$^{1}$}}
\\
$^{1}$Armagh Observatory and Planetarium, College Hill, Armagh, BT61 9DG, UK\\
$^{2}$School of Mathematics and Physics, Queen's University Belfast, Belfast, BT7 1NN, UK\\
$^{3}$School of Physics, Trinity College Dublin, College Green, Dublin 2, Ireland
}
\date{Last updated 2023 December 1 ; in original form 2023 September 29}

\pubyear{2023}

\begin{document}
\label{firstpage}
\pagerange{\pageref{firstpage}--\pageref{lastpage}}
\maketitle

\begin{abstract}
Hydrogen deficient stars include the cool R\,CrB variable (RCBs) and hydrogen-deficient carbon (HdCs) giants through extreme helium stars (EHes) to the very hot helium-rich subdwarfs (He-sdO and O(He) stars) and white dwarfs.
With surfaces rich in helium, nitrogen and carbon, their origins have been identified with the merger of two white dwarfs. 
Using {\it Gaia} to focus on the EHes, we aim to identify progenitor populations and test the evolution models. 
{\it Gaia} DR3 measurements and ground-based radial velocities have been used to compute Galactic orbits using \verb|galpy|.
Each orbit has been classified by population; EHe stars are found in all of the thin disk, thick disk, halo and bulge, as are RCB, HdC and He-sdO stars.
Spectral energy distributions were constructed for all EHes, to provide angular diameters, and hence radii and luminosities. 
The EHes fall into two luminosity groups divided at $L \approx 2500 \, {\rm L_{\odot}}$.
This supports theory for the origin of EHes, and is the strongest confirmation so far in terms of luminosity.
The lower luminosity EHes correspond well with the post-merger evolution of a double helium white dwarf binary. 
Likewise, the higher luminosity EHes match the post-merger evolution of a carbon/oxygen plus helium white dwarf binary. 
In terms of parent populations, current models predict that double white dwarf mergers should occur in all Galactic populations, but  
favour mergers arising from recent star formation (i.e. thin disk), whereas the statistics favour an older epoch (i.e. thick disk). 
\end{abstract}



\section{Introduction}
Extreme helium stars (EHes) are low-mass stars of spectral types A and B \citep{jeffery96a}. 
Their low surface gravities indicate high luminosity-to-mass ratios.
Helium dominates the surface composition, with  hydrogen contributing less than one part per thousand (ppt, by number) and carbon  between 3 and 30 ppt. 
Nitrogen, oxygen and other products of nucleosynthesis are also enhanced \citep{jeffery11a}. 
Their radial-velocities and Galactic distribution suggest an origin in the Galactic bulge \citep{jeffery87}. 
R\,CrB (RCB) and O(He) stars have similar surface chemistries and luminosities but lower and higher surface temperatures respectively \citep{clayton96,reindl14}. The RCB stars represent the most active members of a larger class of cool hydrogen-deficient carbon-rich giants (HdC stars) \citep{crawford23}.       
The surface properties of these three groups are best explained by evolution following the merger of a helium white dwarf with a carbon-oxygen white dwarf \citep{webbink84,saio02}, with their ultimate fate to become a hot hydrogen-deficient white dwarf.  
Naively speaking, an origin in the (old) Galactic bulge should be entirely consistent with the long delay time associated with the orbital decay of a double white dwarf binary via gravitational wave radiation. 

Several EHes have higher surface gravities than the majority and are better understood in terms of evolution following the merger of two helium white dwarfs \citep{saio00}.  
Such stars will evolve toward the helium main-sequence where they are observed as helium-rich sdO stars (He-sdOs) \citep{zhang12a} associated with the Galactic thick disk \citep{martin17b}, and considered younger than the Galactic bulge. 

Binary-star population synthesis (BSPS) calculations suggest other pictures. 
\citet{zhang14} find that double helium white dwarf mergers should come predominantly from star formation more than 4 Gyr ago and 
carbon-oxygen + helium white dwarfs come from star formation between 0.5 and 2 Gyr ago. 
\citet{yu10,yu11} find that the majority of mergers in the present epoch should arise from star formation more than 4 Gyr ago.  
Since actual orbital decay times depend on the white dwarf masses and initial separations, a sufficiently large sample of binaries from a given star-formation epoch that evolve to become double white dwarfs will continue to produce short-lived exotic stars long after their pre-white dwarf evolution is complete. 
Hence - studies of a present day sample of such stars must in fact derive from  a long period of Galactic star formation and reflect a substantial range of actual ages. 

Our main question (1) concerns the progenitor population(s) of EHes; that is, what are their distributions in position and age? 
Secondary questions concern (2) the relationship between EHe stars,  HdC and RCB stars, O(He) stars and He-sdOs, and (3) the validity of the evolution models in terms of EHe masses and luminosities. 
Owing to their considerable heliocentric distances (1 -- 8 kpc), reliable answers to (1) have only become accessible since the 3rd release of data from the {\it Gaia} project \citep{gaia16.mission, gaia23.dr3}. 
In addition, answers to (2) and (3) require additional data for effective temperatures over a wide range of spectral types. 

This paper combines the {\it Gaia} positional and proper motion data and ground-based radial velocity data (\S\,2) to obtain full 6-space positions (\S,3) and population classifications (\S\,4). These data are supplemented with published photometry to build complete spectral energy distributions and hence angular radii and luminosities (\S\,5). 
Masses are inferred from the measured surface gravities, where these are available.  
We present this analysis in detail for the EHes. 
We have also carried out this analysis for comparison samples of related objects including RCB, and helium-rich subdwarfs.
Since similar analyses have been carried out in part or in whole elsewhere \citep[etc.]{ geier20,culpan22,tisserand22}, they are only discussed here in summary form. 

\begin{table*}
\caption[]{Astrometric data for EHe stars from {\it Gaia} including radial velocities, corrected parallax and the computed distance sorted by right ascension.}
\label{t:gaia}
    \begin{tabular}{L{2.2cm} r r r@{$\pm$}l r@{$\pm$}l r@{$\pm$}l r@{$\pm$}l r@{\,}l}
\hline
Star & \multicolumn{1}{c}{$\alpha$}  & \multicolumn{1}{c}{$\delta$}
                & \multicolumn{2}{c}{$\mu_\alpha$} & \multicolumn{2}{c}{$\mu_\delta$} 
                &\multicolumn{2}{c}{$v$}  &\multicolumn{2}{c}{$\pi$} &\multicolumn{2}{c}{Distance}\\
                
 & \multicolumn{2}{c}{deg} & \multicolumn{4}{c}{mas yr$^{\rm -1}$} & \multicolumn{2}{c}{\kmsec} &  \multicolumn{2}{c}{mas} &\multicolumn{2}{c}{kpc} \\
\hline

BD +37 442 & 29.64 & 38.57 & -10.49 & 0.06 & -7.27 & 0.07 & -99.4 & 25.0$^{1}$ & 0.97 & 0.06 & 1.04 & $^{+0.06}_{- 0.07 }$\\
LSS 99 & 103.69 & -10.81 & 0.09 & 0.01 & -0.59 & 0.01 & 109.0 & 3.0$^{2}$ & 0.17 & 0.01 & 5.73 & $^{+0.38}_{- 0.44 }$\\
BD +37 1977 & 141.11 & 36.71 & 10.86 & 0.05 & -11.77 & 0.05 & -80.9 & 9.9$^{1}$ & 0.64 & 0.07 & 1.60 & $^{+0.16}_{- 0.20 }$\\
BD +10 2179 & 159.73 & 10.06 & -10.37 & 0.06 & -4.86 & 0.04 & 155.2 & 2.8$^{2}$ & 0.52 & 0.05 & 1.99 & $^{+0.19}_{- 0.24 }$\\
BX Cir & 210.40 & -66.17 & -8.34 & 0.01 & 2.22 & 0.02 & -83.0 & 2.0$^{3}$ & 0.31 & 0.02 & 3.29 & $^{+0.21}_{- 0.23 }$\\
HD 124448 & 213.74 & -46.29 & -6.54 & 0.04 & -0.05 & 0.03 & -65.0 & 3.0$^{2}$ & 0.65 & 0.03 & 1.56 & $^{+0.07}_{- 0.08 }$\\
PG 1415+492 & 214.26 & 48.96 & 1.1 & 0.03 & -6.55 & 0.03 & 54.0 & 1.0$^{4}$ & 0.37 & 0.03 & 2.76 & $^{+0.21}_{- 0.25 }$\\
CoD --48 10153 & 234.75 & -48.60 & -6.07 & 0.02 & -6.4 & 0.02 & -4.5 & 2.5$^{2}$ & 0.14 & 0.02 & 6.88 & $^{+0.81}_{- 1.03 }$\\
V2205 Oph & 247.15 & -9.33 & -2.27 & 0.03 & 3.06 & 0.02 & -69.3 & 1.0$^{2}$ & 0.14 & 0.03 & 6.85 & $^{+0.95}_{- 1.26 }$\\
V652 Her & 252.02 & 13.26 & -4.24 & 0.04 & 3.76 & 0.03 & 2.0 & 0.5$^{5}$ & 0.66 & 0.05 & 1.55 & $^{+0.11}_{- 0.13 }$\\
V2076 Oph & 265.46 & -17.90 & -6.25 & 0.03 & 2.56 & 0.02 & 76.8 & 2.6$^{2}$ & 0.56 & 0.03 & 1.79 & $^{+0.09}_{- 0.10 }$\\
CoD --46 11775 & 265.64 & -46.98 & -3.79 & 0.03 & -0.95 & 0.02 & -90.7 & 9.3$^{2}$ & 0.21 & 0.03 & 4.91 & $^{+0.62}_{- 0.81 }$\\
LSS 4357 & 266.11 & -19.63 & -3.51 & 0.02 & 1.55 & 0.01 & -99.0 & 10.0$^{2}$ & 0.15 & 0.02 & 6.77 & $^{+0.70}_{- 0.87 }$\\
V2244 Oph & 267.86 & -1.72 & -0.13 & 0.02 & -8.34 & 0.01 & -8.3 & 2.9$^{6}$ & 0.19 & 0.02 & 5.29 & $^{+0.43}_{- 0.51 }$\\
NO Ser & 270.98 & -1.01 & 0.3 & 0.02 & -2.02 & 0.02 & -21.1 & 4.6$^{2}$ & 0.37 & 0.02 & 2.75 & $^{+0.12}_{- 0.13 }$\\
LS IV+6 2 & 271.73 & 6.37 & -3.01 & 0.03 & -8.84 & 0.03 & -24.0 & 4.0$^{2}$ & 0.63 & 0.03 & 1.61 & $^{+0.07}_{- 0.08 }$\\
PV Tel & 275.81 & -56.63 & -1.24 & 0.02 & -7.81 & 0.02 & -170.9 & 2.3$^{2}$ & 0.27 & 0.02 & 3.67 & $^{+0.27}_{- 0.32 }$\\
LSS 5121 & 280.82 & -18.52 & -2.0 & 0.02 & 0.52 & 0.02 & -62.0 & 6.0$^{2}$ & 0.21 & 0.02 & 4.89 & $^{+0.38}_{- 0.45 }$\\
GALEX J184559.8--413827 & 281.50 & -41.64 & 0.24 & 0.04 & 0.1 & 0.03 & -57.6 & 6.1$^{7}$ & 0.18 & 0.04 & 5.56 & $^{+0.90}_{- 1.27 }$\\
LS IV --14 109 & 284.91 & -14.44 & -0.36 & 0.02 & -0.43 & 0.02 & -7.3 & 2.7$^{8}$ & 0.25 & 0.02 & 3.96 & $^{+0.29}_{- 0.34 }$\\
GALEX J191049.5--441713 & 287.71 & -44.29 & -2.7 & 0.05 & -2.93 & 0.04 & 1.0 & 3.0$^{9}$ & 0.70 & 0.05 & 1.45 & $^{+0.09}_{- 0.11 }$\\
V1920 Cyg & 296.32 & 33.97 & 0.11 & 0.02 & -6.24 & 0.02 & -89.9 & 1.9$^{2}$ & 0.23 & 0.02 & 4.39 & $^{+0.29}_{- 0.33 }$\\
EC 19529--4430 & 299.13 & -44.37 & -6.8 & 0.03 & -13.22 & 0.03 & 5.0 & 3.0$^{9}$ & 0.21 & 0.04 & 4.77 & $^{+0.71}_{- 0.98 }$\\
EC 20111--6902 & 304.04 & -68.89 & 2.65 & 0.06 & -1.72 & 0.08 & -81.0 & 3.0$^{9}$ & 0.19 & 0.08 & 5.08 & $^{+1.27}_{- 1.96 }$\\
EC 20236--5703 & 306.90 & -56.90 & 1.94 & 0.04 & -13.34 & 0.04 & -74.0 & 3.0$^{9}$ & 0.26 & 0.04 & 3.94 & $^{+0.50}_{- 0.67 }$\\
BPS CS 22940--0009 & 307.58 & -59.84 & 2.53 & 0.03 & -10.79 & 0.03 & 32.7 & 2.0$^{4}$ & 0.47 & 0.03 & 2.16 & $^{+0.15}_{- 0.17 }$\\
FQ Aqr & 312.84 & 2.31 & -0.62 & 0.02 & -6.18 & 0.02 & 23.7 & 1.4$^{6}$ & 0.21 & 0.02 & 4.84 & $^{+0.39}_{- 0.46 }$\\

\hline
\multicolumn{13}{p{14cm}}{1.\citet{deBruijne2012}, 2.\citet{jeffery87}, 3.\citet{kilkenny99}, 4.\citet{martin17a}, 5.\citet{jeffery15a}, 6.\citet{jeffery01c}, 7.\citet{kawka15}, 8.\citet{lawson93}, 9.\citet{jeffery21a}}
\end{tabular}\\
\end{table*}

\section{Observations}

The complete dataset for analysis contains the full 6-space position for every known member of the different classes of hydrogen-deficient stars together with spectroscopic and photometric measurements of effective temperature, surface gravity and interstellar extinction. 
The EHe sample was based on compilations by \citet{jeffery96,clayton96} and \citet{martin19.phd}.
\footnote{DY\,Cen has been excluded from the EHe sample on the basis of its high hydrogen content and very rapid evolution \citep{jeffery20a}. HD 144941 has been excluded as possibly the most extreme member of the magnetic massive helium-strong stars \citep{przybilla21a}.} 
In addition, we have included new EHes discovered as part of the SALT survey of chemically peculiar subdwarfs \citep{jeffery21a}. 
The primary criteria for inclusion are that each star should be included in {\it Gaia} Data Release 3 (DR3) and should have a known radial velocity. 

The majority of EHe radial velocities come from \citet{jeffery87} and \citet{jeffery21a}. 
The former found no significant variability amongst all EHes surveyed. 
With no evidence that any are binaries, \citet{jeffery87} argued that all must have been binaries at some time in the past. 
Where repeat observations have been obtained, none of the stars classified by \citet{jeffery21a} and identified here as EHes have yet been found to be binaries. 
A number are known to pulsate with irregular small-amplitude radial-velocity variations \citep{walker85,lawson93,jeffery01c}, for which we adopt the mean velocity, or with regular large-amplitude radial-velocity variations  \citep{lynasgray84,kilkenny99}, for which the system velocity is known precisely.
We take the space velocities to be secure to within the errors shown.
For the EHes, the typical error is $\sim \pm3 \kmsec$ and for the secondary sample, the typical error is $\sim \pm17 \kmsec$.

For the EHe sample, an additional criterion was the availability of measurements for effective temperatures ($T_{\rm eff}$) and surface gravities ($\log g$). These measurements along with the extinction coefficients ($E_{\rm (B-V)}$) are presented in section \ref{luminosities&radii}. 
Table\,\ref{t:gaia} shows the {\it Gaia} data (position, proper motion and parallax) and the radial velocity for the 27 stars in the EHe sample.

We studied the kinematics and the Galactic orbits of these stars and derived the angular diameters and luminosities from their spectral energy distributions. To compare this sample with the other groups mentioned above, we compiled a secondary sample containing HdC, RCB, He-sdO, He-sdOB, and He-sdB stars based on \citet{clayton12,reindl14,geier20,tisserand20,culpan22,tisserand22} and \citet{crawford23}. 
A list of these stars is provided in Appendix\,\ref{a:sample}.

\begin{figure}
\includegraphics[width=0.53\textwidth]{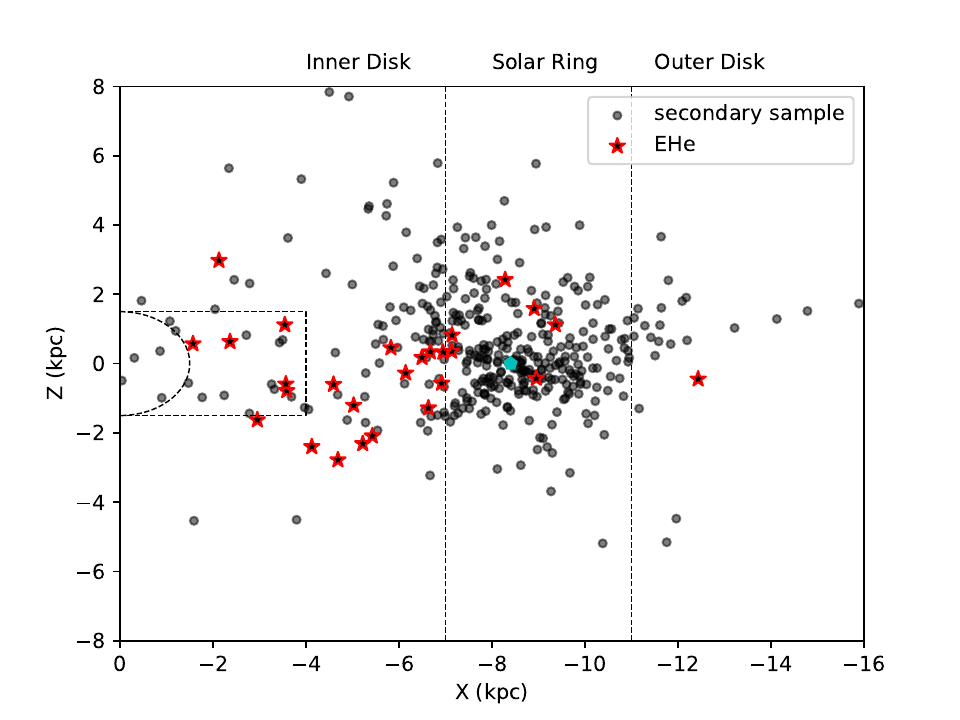}
\caption{Distribution of sample in the Galactic $X - Z$ plane. The red stars denote the primary sample and the grey dots denote the secondary sample. The position of the Sun is denoted by a filled cyan pentagon. The dashed ellipse shows the estimates of the inner bulge (R $\sim$ 1.5 kpc) and the dashed rectangle shows the estimates of the bar (X = $\pm$4 kpc, Y = $\pm$ 1.5 kpc, Z = $\pm$ 1.5 kpc) \citep{galreview16}. }
\label{XZ}
\end{figure}


\begin{figure*}
    \centering
    \includegraphics[width=0.8\textwidth]{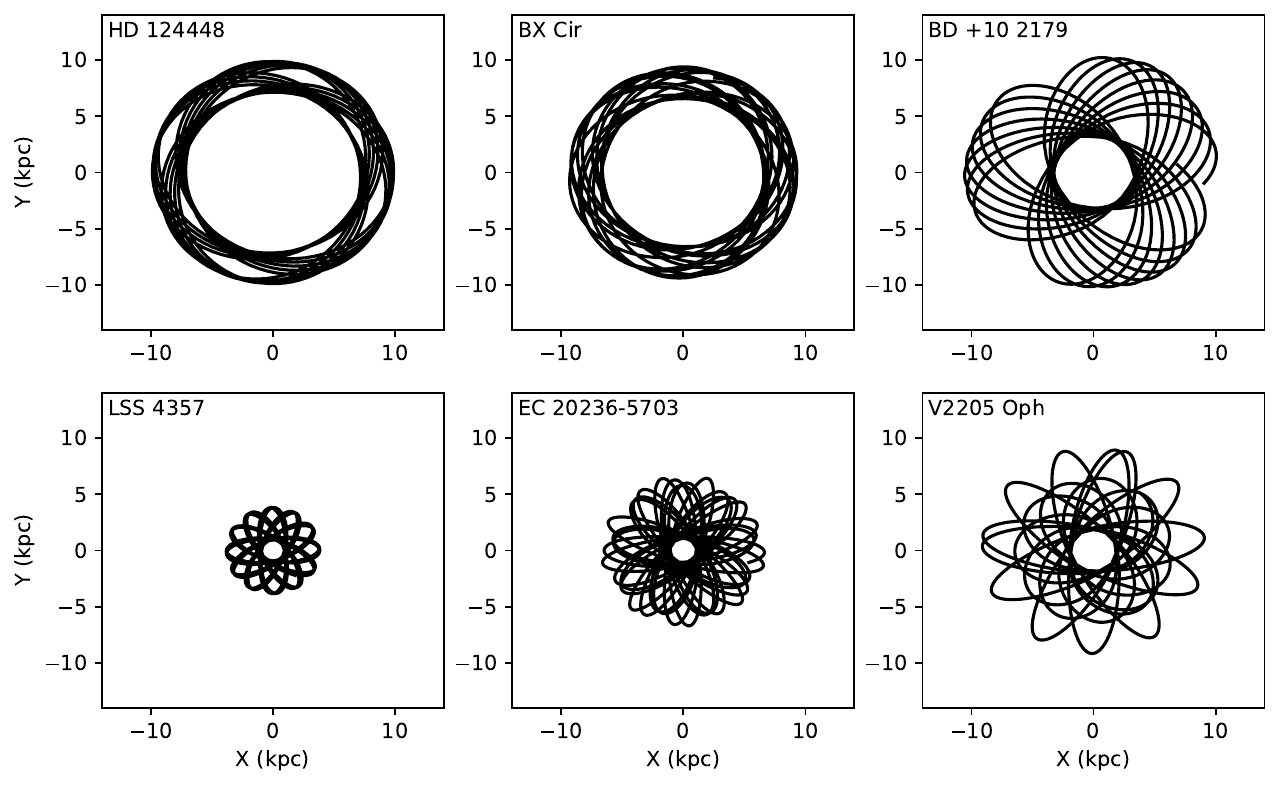}
    \includegraphics[width=0.8\textwidth]{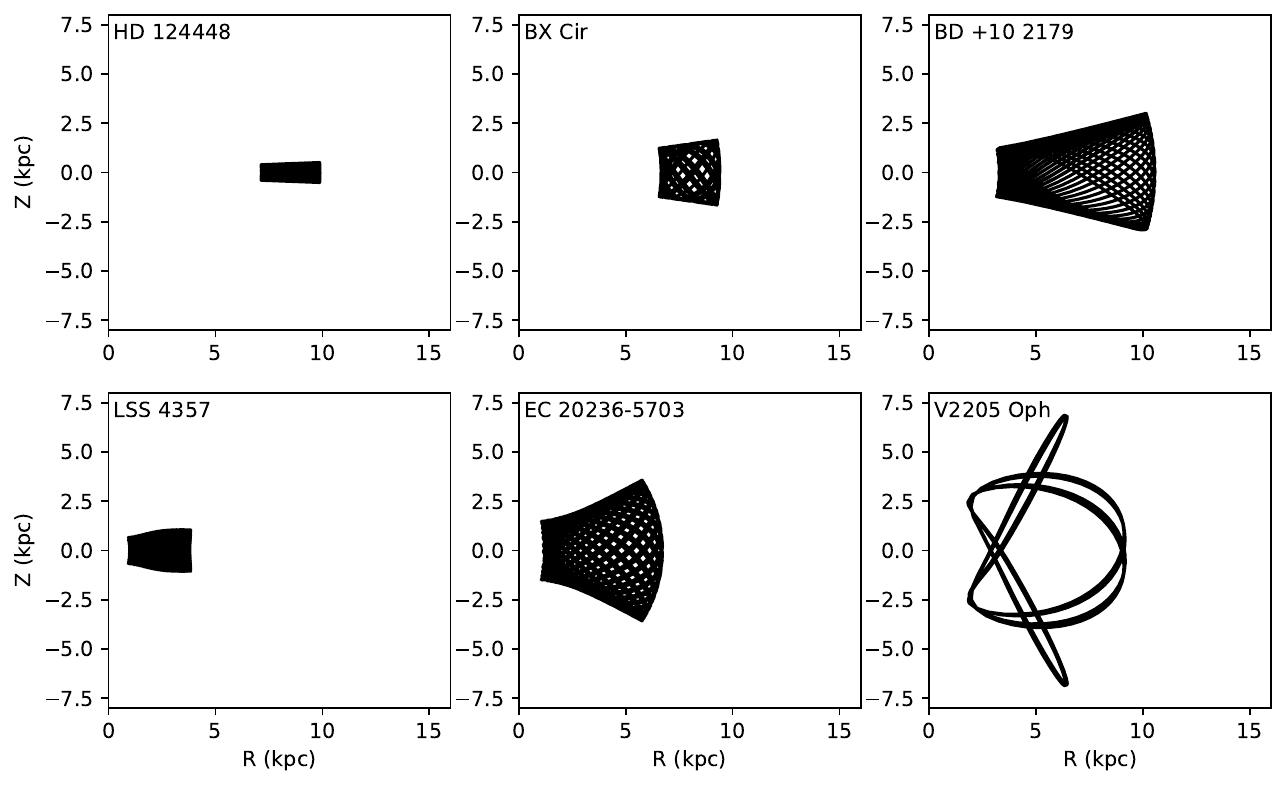}
    \caption{\texttt{galpy} orbits for 6 representative EHe stars computed for 3 Gyrs from their current positions. 
    The top panels show motion in the $\rm X-Y$ plane, the bottom panels show motion in the $\rm R-Z$ plane, with the Galactic centre being at the origin. 
    The panels illustrate thin-disk (HD\,124448), thick disk (BX\,Cir), halo (BD\,+10\,2179), bulge (LSS\,4357), halo (EC\,20236--5703), and halo (V2205\,Oph) orbits.} 
    \label{galpy}
\end{figure*}

\begin{figure}
\includegraphics[width=0.5\textwidth]{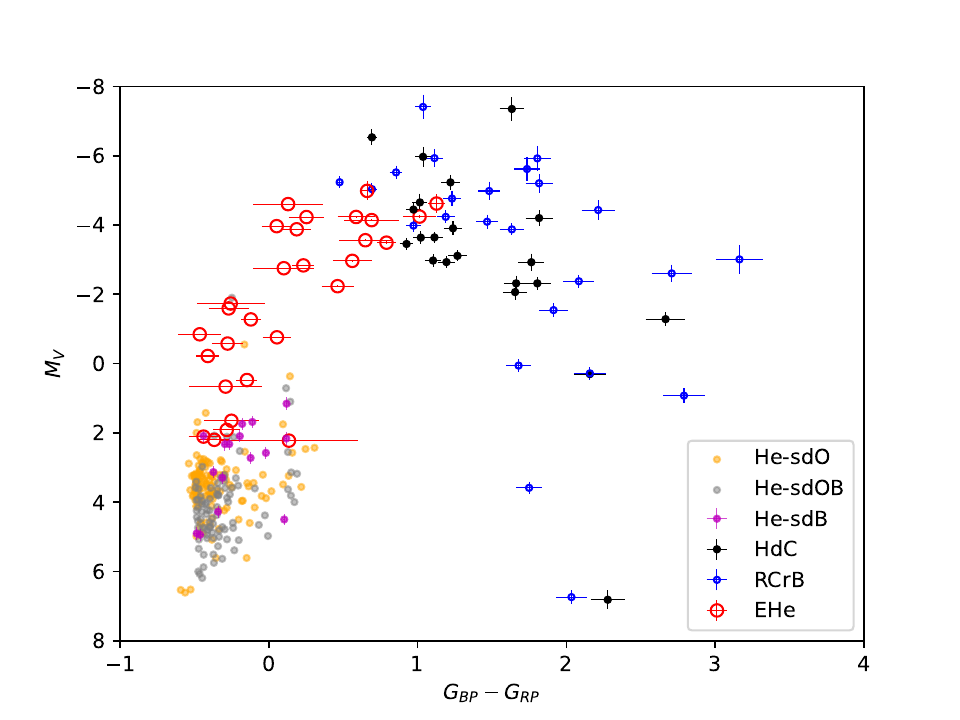}
\caption{Colour magnitude diagram of the whole sample obtained from the {\it Gaia} magnitude and colours as given in S. \ref{Gaia_colors}.}
\label{CMD}
\end{figure}

\section{Analysis}

\subsection{Distances}

{\it Gaia} DR3 parallaxes have been used to obtain distances. 
The zero points of the parallax measurements have been corrected as described by \citet{lindegren21} and the corrected parallax is given in Table \ref{t:gaia}. 

Inverting a parallax to give a distance is only appropriate when measurement errors are negligible. 
As parallax measurements always contain significant errors, determining distance from parallax becomes an inference problem.
Small absolute errors can translate into large errors in distance \citep{bailerjones15,luri18,bailerjones21}.
Once the fractional parallax error $\sigma/\pi \geq 0.2$,  a simple inversion becomes noisy and gives an incorrect error estimate.
This can be mitigated by the use of a properly normalised prior. 
Here an Exponentially Decreasing Space Density Prior is adopted, 
with  one free parameter, the length scale, fixed at $1.35$\,kpc 
\citep{luri18}.
Even though a Bayesian approach can give distance estimates from even a negative parallax measurement, the prior will always dominate when the fractional error in parallax  is high.
Three stars in the primary sample have fractional parallax error $\sigma/\pi \geq 0.2$: EC 20111--6902 (0.42), GALEX J184559.8--413827 (0.22) and V2205 Oph (0.21). 
We note that although these errors are significant, they are substantially improved over {\it Gaia} DR2. 
In particular, V2205\,Oph had $\sigma/\pi \geq 0.5$ in {\it Gaia} DR2.

Distance measurements obtained for the stars in the primary sample are reported in Table \ref{t:gaia}, where the errors correspond to the 68.27\% confidence intervals. 

\subsection{Galactic and orbital parameters}
Figure \ref{XZ} shows the positions of the stars projected onto the Galactic $X - Z$ plane, where $Z$ is the distance from the Galactic plane, and $X$ from the Galactic Centre (GC).
We define the inner disk $R < 7$\,kpc, the Solar ring  $7 < R < 11$\,kpc and the outer disk $R>11$\,kpc from the GC \citep{galreview16}.  
Estimates of the extent of the bulge and/or bar are also shown.
Most of the stars in our sample lie between the Sun and the GC. Seven EHe stars are situated within or close to the bulge or bar. This highlights the capabilities of {\it Gaia} in separating out sources that are close to the Galactic bulge. This figure also highlights that our sample is subject to Malmquist bias, or the preferential detection at large distances of intrinsically brighter objects. The question this observation raises is whether this constraint causes any bias in our understanding of group kinematics and further inferences.

Using the observed values of right ascension ($\alpha$), declination ($\delta$), distance, proper motion ($\mu_{\alpha}$ and $\mu_{\delta}$) and radial velocity ($v$), the Galactic velocity components were calculated following the method outlined in \citet{randall15}. 
The left-handed system for the velocity components is used here, where $U$ is the Galactic radial velocity, positive towards the Galactic Centre, $V$ is the Galactic rotational velocity in the direction of the Galactic rotation and $W$ is the component positive towards the North Galactic Pole. 
This calculation assumes the distance of the Sun from the Galactic Centre to be $8.25\pm0.33$ kpc \citep{gravity21}, its motion relative to the LSR to have components $(v_x, v_y, v_z)=(9.5,17.3,8.56)$ km $\rm s^{-1}$ and the velocity of the LSR to be $V_{\rm LSR}=233.4\pm1.5$ km $\rm s^{-1}$ \citep{drimmel18,reid20,gravity21}.

Orbits for individual stars (Fig.\,\ref{galpy}) have been calculated using \verb|galpy|\footnote{http://github.com/jobovy/galpy}, a python package for Galactic-dynamic calculations \citep{bovy15}. The orbits were computed using the potential MWPotential 2014; this model is fit to dynamical data of the Milky Way. Although this is not the best possible current model, it was chosen as it gives a realistic model of the Milky Way’s gravitational potential that is simple and easy to use. It consists of a bulge modelled as a power-law density profile that is exponentially cut off with a power-law exponent of –1.8 and a cut-off radius of 1.9 kpc, a Miyamoto–Nagai Potential disc \citep{miyamoto75} and a dark-matter halo described by a Navarro–Frenk–White potential \citep{navarro96}.
These orbits when integrated over 3\,Gyrs give values for the maximum distance from the GC (apocentre $R_{\rm a}$), minimum distance to the GC (pericentre $R_{\rm p})$, maximum vertical amplitude ($Z_{\rm max}$) and the z-component of the angular momentum ($J_z$). The quantities $R_{\rm a}$ and $R_{\rm p}$ are obtained over a revolution of $2\pi$ radians, measured on the Galactic plane. From these distances, the orbital eccentricity can be computed as
\begin{equation}
    e = \frac{R_{\rm a} - R_{\rm p}}{R_{\rm a} + R_{\rm p}}.
\end{equation}

Using \verb|galpy|, 100 orbits were computed for each star, with starting values distributed randomly over the range of errors associated with each one so as to obtain a sample of likely orbits. From this sample, the orbit with the highest likelihood (mode) was selected with the rest being used to compute the statistical errors (standard deviation) on the orbital elements. These quantities are shown in Table \ref{A2}, along with the stars' Galactic velocities.

\subsection{Colour-magnitude diagram}
\label{Gaia_colors}
When combined with the extinction coefficient ($E_{\rm B - V}$), {\it Gaia} distances and colours provide us with absolute magnitudes $M_{\rm V}$. These can be defined in terms of the apparent {\it Gaia} magnitude $G$ and colours $G_{\rm BP}-G_{\rm RP}$ using the Johnson-Cousins relation from \citet{riello21}: 

\begin{dmath}
V = G + 0.02704 - 0.01424(G_{\rm BP} - G_{\rm RP}) + 0.2156(G_{\rm BP} - G_{\rm RP})^2 \\ - 0.01426(G_{\rm BP} - G_{\rm RP})^3,
\end{dmath}
\begin{equation}
\label{eq:mv}
M _{\rm V} = V - 5(\log_{10} d - 1)- 3.1 E _{\rm B - V},
\end{equation}
where $V$ is the apparent visual magnitude and $d$ is the distance in parsecs. 

Since Eqn (2) has been derived for stars with hydrogen-rich atmospheres, we compared the transformed V magnitude with V magnitudes from SIMBAD. 
The latter are, on average, $\approx 0.02$ mag larger for $G_{\rm BP}-G_{\rm RP} > 0$ decreasing to $\approx 0.05$ mag smaller for $G_{\rm BP}-G_{\rm RP} \approx-0.4 $, with a mean scatter  of $\approx\pm0.02$ mag (1$\sigma$). 
The small trend is consistent with a reduced contribution of the Balmer series to the EHe star colours.  
The transformed EHe V magnitudes may be improved by applying the correction 
\begin{equation} 
\delta = 0.00759 + 0.11452 (G_{\rm BP} - G_{\rm RP}) - 0.11589 (G_{\rm BP} - G_{\rm RP})^2.   
\end{equation}


The colour-magnitude diagram of the whole sample plotted from the {\it Gaia} colours\footnote{Author’s note in proof: Fig.\,\ref{CMD} is based on mean Gaia magnitudes. For RCB and HdC stars observed during a decline, mean Gaia magnitudes do not reflect the luminosity at maximum light and here applies to RCB and HdC stars with $M_{\rm V} > -3$.} is shown in Fig.\,\ref{CMD}. This shows a progression from the cool HdC stars and RCB stars to the hotter EHe and He-SdO stars. There is an overlap between the coolest EHe stars and the hottest HdCs.

\begin{table*}
\caption[]{Orbital parameters and Galactic velocities of Extreme Helium stars  sorted as in Table \ref{t:gaia}. The last column shows the population classification where TH = thin disk, TK = thick disk, H = halo and B = bulge. $f = p({\rm thick})/p({\rm thin})$ shows the likelihood of a star belonging to the thick or thin disk. The stars for which the orbits were classified by inspection have been marked with a *.}
\label{A2}
\setlength{\tabcolsep}{4pt}
\tabulinesep=1.5mm
\begin{tabular}{ L{2cm} r@{\,}l r@{\,}l r@{\,}l r@{$\pm$}l r@{$\pm$}l r@{$\pm$}l r@{$\pm$}l r@{$\pm$}l r c}
\hline
Star & \multicolumn{2}{c}{$U$} & \multicolumn{2}{c}{$V$} & \multicolumn{2}{c}{$W$} & 
    \multicolumn{2}{c}{$J_z$} &
    \multicolumn{2}{c}{$e$} & \multicolumn{2}{c}{R$_{a}$} &
    \multicolumn{2}{c}{R$_p$} & \multicolumn{2}{c}{Z$_{\rm max}$} & $\log(f)$ &pop \\
  & \multicolumn{2}{c}{\kmsec} & \multicolumn{2}{c}{\kmsec} & \multicolumn{2}{c}{\kmsec} & \multicolumn{2}{c}{kpc \kmsec} & \multicolumn{2}{c}{ }   &
    \multicolumn{2}{c}{kpc} & \multicolumn{2}{c}{kpc}  &
    \multicolumn{2}{c}{kpc} &  &\\
\hline
BD +37 442 & $-98.1$ & $^{+10.5}_{- 10.6 }$ & $210.1$ & $^{+8.5}_{- 8.4 }$ & $0.9$ & $^{+5.7}_{- 5.6 }$ & $1971.9$ & $75.3$ & $0.33$ & $0.04$ & $12.03$ & $0.19$ & $6.05$ & $0.44$ & $0.47$ & $0.03$ & $-0.82$&TH\\
LSS 99 & $10.0$ & $^{+1.6}_{- 1.8 }$ & $175.3$ & $^{+0.8}_{- 0.8 }$ & $-5.0$ & $^{+0.2}_{- 0.2 }$ & $2352.5$ & $31.3$ & $0.25$ & $0.00$ & $13.19$ & $0.20$ & $7.86$ & $0.13$ & $0.42$ & $0.02$ & $-0.85$&TH\\
BD +37 1977 & $-20.4$ & $^{+4.3}_{- 4.5 }$ & $114.4$ & $^{+4.3}_{- 5.1 }$ & $-20.8$ & $^{+4.3}_{- 4.4 }$ & $1646.3$ & $44.1$ & $0.48$ & $0.02$ & $12.89$ & $0.28$ & $4.52$ & $0.18$ & $1.68$ & $0.20$ & -&H\\
BD +10 $2179^{*}$ & $-79.6$ & $^{+3.2}_{- 4.4 }$ & $189.9$ & $^{+0.7}_{- 0.7 }$ & $110.2$ & $^{+1.3}_{- 1.6 }$ & $1161.0$ & $26.1$ & $0.53$ & $0.02$ & $10.59$ & $0.15$ & $3.26$ & $0.17$ & $3.01$ & $0.06$ & -&H\\
BX Cir$^{*}$ & $-130.2$ & $^{+3.4}_{- 3.4 }$ & $280.7$ & $^{+2.1}_{- 1.8 }$ & $114.7$ & $^{+1.9}_{- 1.9 }$ & $1827.1$ & $20.3$ & $0.17$ & $0.01$ & $9.42$ & $0.14$ & $6.65$ & $0.03$ & $1.66$ & $0.06$ & $0.91$&TK\\
HD 124448 & $-34.9$ & $^{+1.5}_{- 1.4 }$ & $292.6$ & $^{+1.2}_{- 1.2 }$ & $22.0$ & $^{+0.6}_{- 0.6 }$ & $2003.4$ & $16.7$ & $0.16$ & $0.01$ & $9.89$ & $0.17$ & $7.14$ & $0.03$ & $0.51$ & $0.01$ & $-0.8$&TH\\
PG 1415+492 & $-27.1$ & $^{+0.6}_{- 0.8 }$ & $230.3$ & $^{+0.9}_{- 1.2 }$ & $75.4$ & $^{+0.7}_{- 0.7 }$ & $2094.5$ & $7.8$ & $0.16$ & $0.00$ & $11.90$ & $0.08$ & $8.67$ & $0.03$ & $4.60$ & $0.19$ & $ 13.57$&TK\\
CoD --48 10153 & $-140.9$ & $^{+3.9}_{- 0.9 }$ & $0.9$ & $^{+11.8}_{- 15.9 }$ & $7.5$ & $^{+0.5}_{- 0.5 }$ & $494.0$ & $19.9$ & $0.51$ & $0.02$ & $4.59$ & $0.03$ & $1.50$ & $0.09$ & $0.85$ & $0.09$ & -&H\\
V2205 Oph & $111.3$ & $^{+20.1}_{- 10.8 }$ & $299.7$ & $^{+6.6}_{- 1.1 }$ & $103.5$ & $^{+5.2}_{- 6.9 }$ & $563.8$ & $172.2$ & $0.57$ & $0.07$ & $9.39$ & $0.53$ & $2.59$ & $0.36$ & $6.87$ & $1.47$ & -&H\\
V652 Her & $4.8$ & $^{+0.5}_{- 0.6 }$ & $269.4$ & $^{+0.3}_{- 0.5 }$ & $50.4$ & $^{+0.8}_{- 1.1 }$ & $1847.5$ & $12.8$ & $0.14$ & $0.01$ & $9.05$ & $0.02$ & $6.87$ & $0.08$ & $1.31$ & $0.06$ & $0.21$&TK\\
V2076 Oph & $34.3$ & $^{+1.6}_{- 1.5 }$ & $251.8$ & $^{+0.4}_{- 0.4 }$ & $66.2$ & $^{+1.0}_{- 1.1 }$ & $1525.8$ & $16.4$ & $0.24$ & $0.01$ & $8.25$ & $0.07$ & $5.04$ & $0.06$ & $0.86$ & $0.03$ & $-0.47$&TH\\
CoD --46 11775 & $-58.2$ & $^{+8.6}_{- 9.6 }$ & $235.2$ & $^{+5.0}_{- 4.2 }$ & $98.8$ & $^{+2.7}_{- 4.0 }$ & $910.6$ & $98.3$ & $0.02$ & $0.02$ & $4.20$ & $0.36$ & $4.00$ & $0.34$ & $1.17$ & $0.06$ & -&B\\
LSS 4357 & $135.1$ & $^{+18.4}_{- 14.9 }$ & $234.4$ & $^{+13.3}_{- 6.0 }$ & $142.6$ & $^{+4.3}_{- 5.6 }$ & $297.6$ & $115.8$ & $0.61$ & $0.10$ & $3.98$ & $0.22$ & $0.98$ & $0.35$ & $1.06$ & $0.27$ & -&B\\
V2244 Oph & $-39.3$ & $^{+1.2}_{- 1.4 }$ & $128.4$ & $^{+1.6}_{- 2.2 }$ & $-85.8$ & $^{+2.2}_{- 2.7 }$ & $545.7$ & $31.7$ & $0.40$ & $0.03$ & $4.73$ & $0.05$ & $2.04$ & $0.13$ & $1.71$ & $0.08$ & - &B\\
NO Ser & $45.6$ & $^{+2.3}_{- 2.3 }$ & $217.2$ & $^{+1.8}_{- 1.7 }$ & $-11.3$ & $^{+0.6}_{- 0.6 }$ & $1362.2$ & $19.7$ & $0.15$ & $0.01$ & $6.51$ & $0.05$ & $4.85$ & $0.10$ & $0.55$ & $0.02$ & $-0.77$&TH\\
LS IV+6 2 & $-25.5$ & $^{+1.7}_{- 1.8 }$ & $195.1$ & $^{+1.5}_{- 1.5 }$ & $-1.6$ & $^{+0.5}_{- 0.5 }$ & $1331.3$ & $20.5$ & $0.25$ & $0.01$ & $7.17$ & $0.03$ & $4.29$ & $0.10$ & $0.38$ & $0.01$ & $-0.87$&TH\\
PV Tel & $135.9$ & $^{+1.1}_{- 1.0 }$ & $236.4$ & $^{+1.2}_{- 1.4 }$ & $38.0$ & $^{+0.7}_{- 0.8 }$ & $1256.8$ & $40.2$ & $0.40$ & $0.00$ & $8.94$ & $0.27$ & $3.84$ & $0.09$ & $2.16$ & $0.04$ & $2.23$&TK\\
LSS 5121 & $93.5$ & $^{+4.0}_{- 4.3 }$ & $223.2$ & $^{+2.3}_{- 2.1 }$ & $68.6$ & $^{+1.3}_{- 1.5 }$ & $802.2$ & $53.4$ & $0.32$ & $0.02$ & $5.15$ & $0.15$ & $2.65$ & $0.18$ & $0.98$ & $0.04$ & -&B\\
GALEX J184559.8--413827 & $1.9$ & $^{+8.3}_{- 6.4 }$ & $264.3$ & $^{+1.2}_{- 1.1 }$ & $19.7$ & $^{+1.4}_{- 1.3 }$ & $847.1$ & $155.5$ & $0.22$ & $0.05$ & $5.28$ & $0.26$ & $3.35$ & $0.54$ & $2.12$ & $0.34$ & $2.1$&TK\\
LS IV --14 109 & $59.7$ & $^{+2.2}_{- 2.7 }$ & $233.6$ & $^{+1.1}_{- 1.2 }$ & $13.1$ & $^{+0.3}_{- 0.3 }$ & $1160.5$ & $39.4$ & $0.17$ & $0.01$ & $5.78$ & $0.09$ & $4.08$ & $0.17$ & $0.64$ & $0.03$ & $-0.69$&TH\\
GALEX J191049.5--441713 & $-23.7$ & $^{+1.6}_{- 1.6 }$ & $228.7$ & $^{+0.5}_{- 0.6 }$ & $20.5$ & $^{+0.7}_{- 0.7 }$ & $1617.5$ & $17.4$ & $0.09$ & $0.01$ & $7.26$ & $0.05$ & $6.11$ & $0.09$ & $0.58$ & $0.02$ & $-0.74$&TH\\
V1920 Cyg & $-0.9$ & $^{+0.5}_{- 0.6 }$ & $155.2$ & $^{+1.2}_{- 1.2 }$ & $-64.6$ & $^{+1.3}_{- 1.5 }$ & $1271.5$ & $15.0$ & $0.33$ & $0.01$ & $7.99$ & $0.03$ & $4.01$ & $0.09$ & $1.48$ & $0.10$ & $ 0.53$&TK\\
EC 19529--4430 & $-61.3$ & $^{+1.5}_{- 2.2 }$ & $-61.1$ & $^{+13.0}_{- 18.2 }$ & $98.3$ & $^{+4.0}_{- 5.4 }$ & $-314.5$ & $117.4$ & $0.64$ & $0.15$ & $5.20$ & $0.29$ & $1.13$ & $0.55$ & $2.99$ & $0.22$ & -&H\\
EC 20111--6902 & $-40.8$ & $^{+11.9}_{- 13.7 }$ & $279.8$ & $^{+3.4}_{- 2.1 }$ & $8.0$ & $^{+3.3}_{- 5.1 }$ & $1523.0$ & $138.9$ & $0.26$ & $0.02$ & $9.91$ & $0.14$ & $5.86$ & $0.29$ & $4.26$ & $0.92$ & $11.47$&TK\\
EC 20236--5703 & $127.7$ & $^{+4.6}_{- 5.7 }$ & $67.1$ & $^{+6.0}_{- 9.1 }$ & $36.2$ & $^{+1.1}_{- 1.2 }$ & $368.5$ & $101.0$ & $0.71$ & $0.05$ & $6.79$ & $0.06$ & $1.15$ & $0.27$ & $3.57$ & $0.42$ & -&H\\
BPS CS 22940--0009 & $-0.5$ & $^{+1.2}_{- 1.3 }$ & $142.4$ & $^{+2.1}_{- 2.2 }$ & $-20.5$ & $^{+0.7}_{- 0.7 }$ & $982.0$ & $39.3$ & $0.40$ & $0.02$ & $6.94$ & $0.06$ & $2.96$ & $0.14$ & $1.32$ & $0.06$ & $0.27$&TK\\
FQ Aqr & $-3.9$ & $^{+0.4}_{- 0.4 }$ & $198.7$ & $^{+1.2}_{- 1.4 }$ & $-56.3$ & $^{+1.2}_{- 1.6 }$ & $1290.5$ & $22.3$ & $0.13$ & $0.01$ & $6.88$ & $0.01$ & $5.28$ & $0.08$ & $2.47$ & $0.14$ & $3.2$&TK\\

\hline
\end{tabular}
\end{table*}

\begin{figure}
\includegraphics[width=0.5\textwidth]{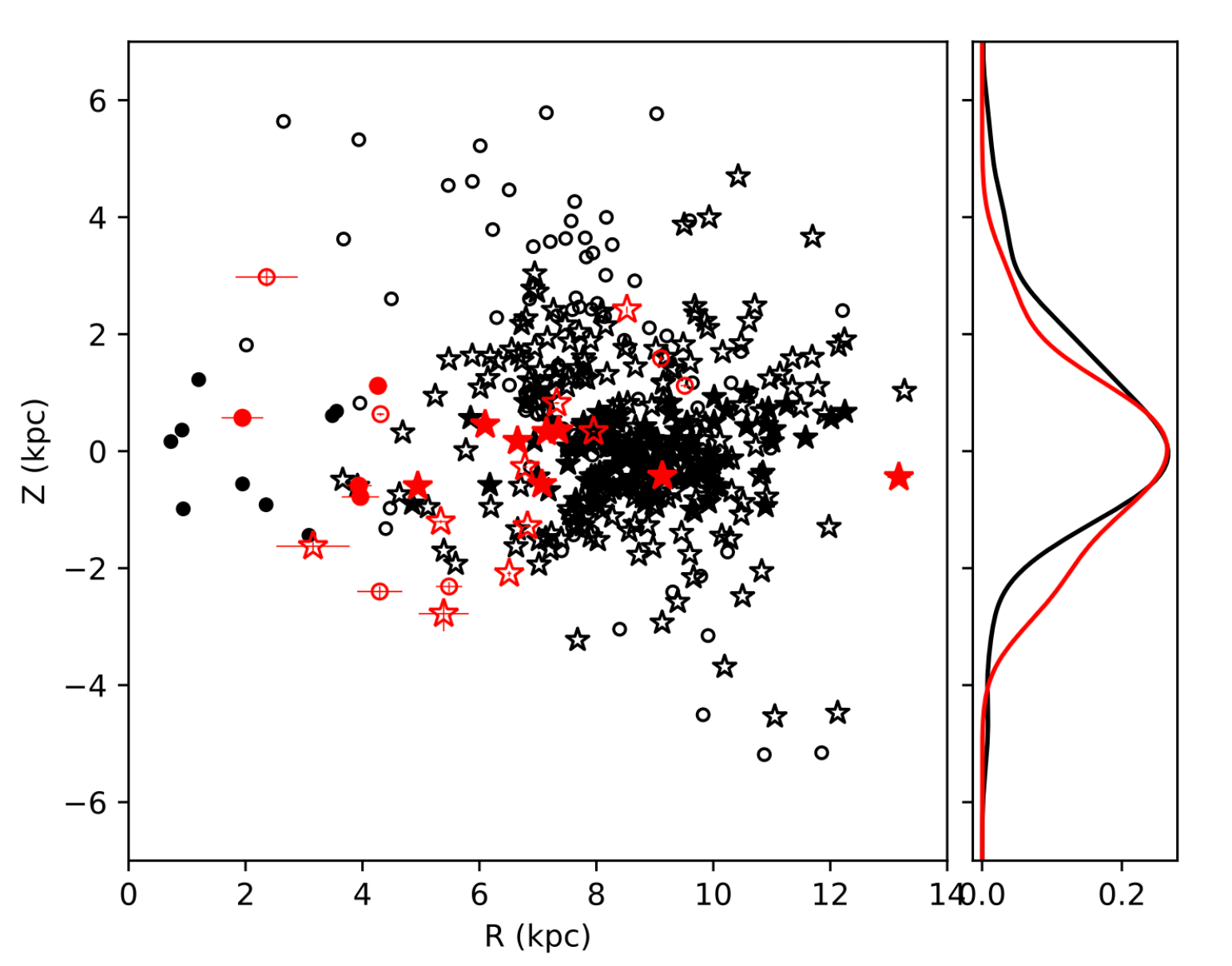}
\caption{Distribution of the sample in the Z plane plotted as a function of the galactocentric distance. Red symbols indicate the primary sample of EHes and the black symbols denote the secondary sample. The sample has been classified into the disc and spherical components as given in section \ref{kinematics}. The circles represent the spherical components with the filled circles denoting the bulge and the open circles denoting the halo stars. The stars represent the disk components with the filled stars denoting the thin disk and the open circles denoting the thick disk stars.}
\label{RZ}
\end{figure}
\begin{figure}
\includegraphics[width=0.51\textwidth]{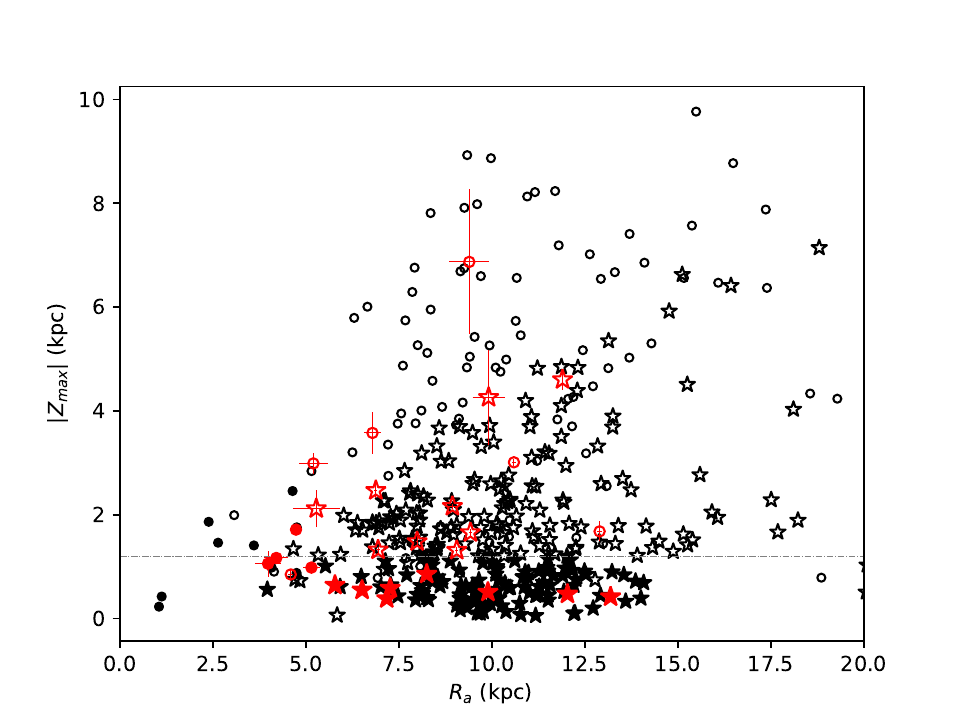}
\caption{The maximum distance in the Z  direction ($\rm Z_{max}$) is plotted as a function of the maximum distance of the star from the Galactic center ($\rm R_{a}$). The symbols have the same meaning as defined in Fig. \ref{RZ}. The dashed line represents the thin disk cut-off height of 1.2 kpc set to distinguish between the thick and thin disk.}
\label{Rapzmax}
\end{figure}
\begin{figure}
\includegraphics[width=0.5\textwidth]{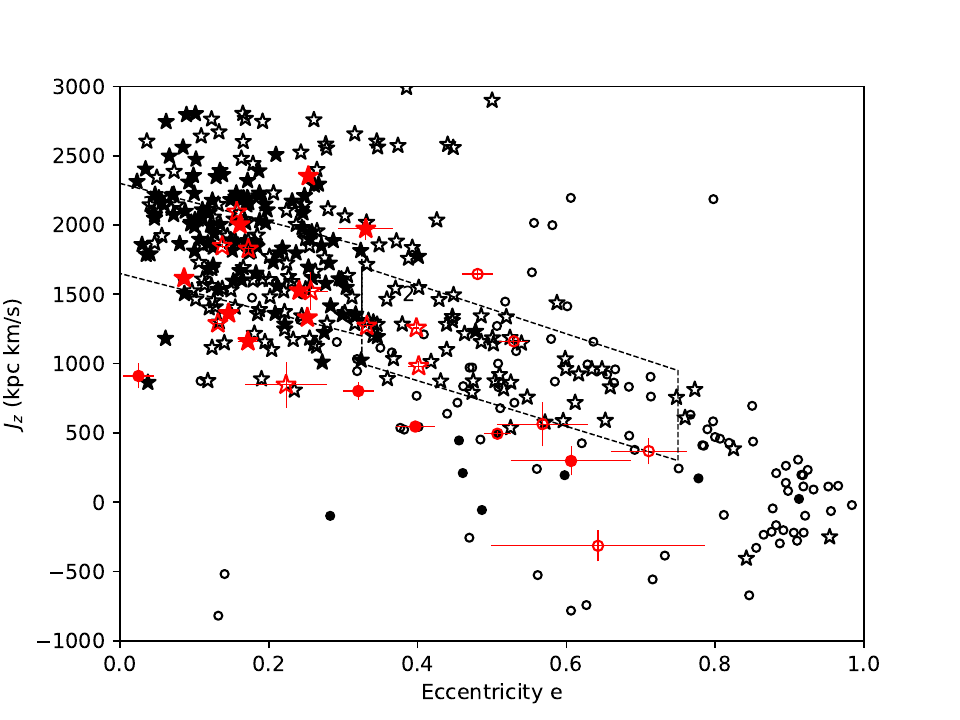}
\caption{The z component of angular momentum plotted as a function of eccentricity. The symbols have the same meaning as defined in Fig. \ref{RZ}. The dashed regions indicate estimates for the thin and thick disk stars from \citet{pauli03}.}
\label{eJ}
\end{figure}
\begin{figure}
\includegraphics[width=0.45\textwidth]{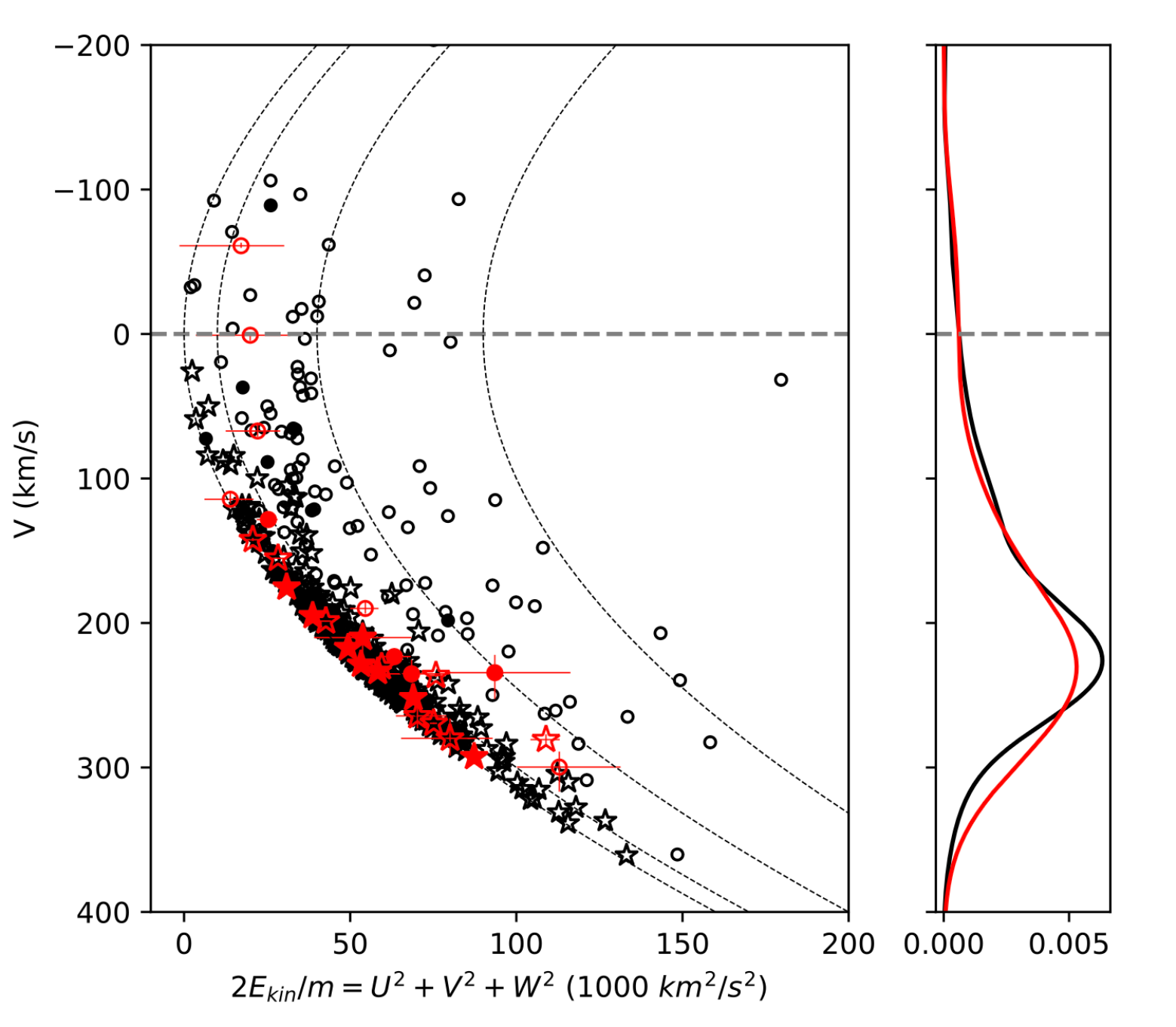}
\caption{Galactic rotational velocity plotted against the total kinetic energy. The symbols have the same meaning as defined in Fig. \ref{RZ}. The parabolic curves denote line of equal velocity ($V_{\bot} =(U^{2} + W^{2} )^{1/2}$) at $0, 100, 300$ and $400 \kmsec$ respectively. The histogram with a binsize $=\,20\kmsec$ depicts the Galactic rotational velocities of the whole sample. The red and black bins refer to their respective samples. The grey dashed line represents $V = 0$ with the stars above it showing retrograde motion and the stars below it showing prograde motion. }
\label{kin}
\end{figure}
\begin{figure}
\includegraphics[width=0.53\textwidth]{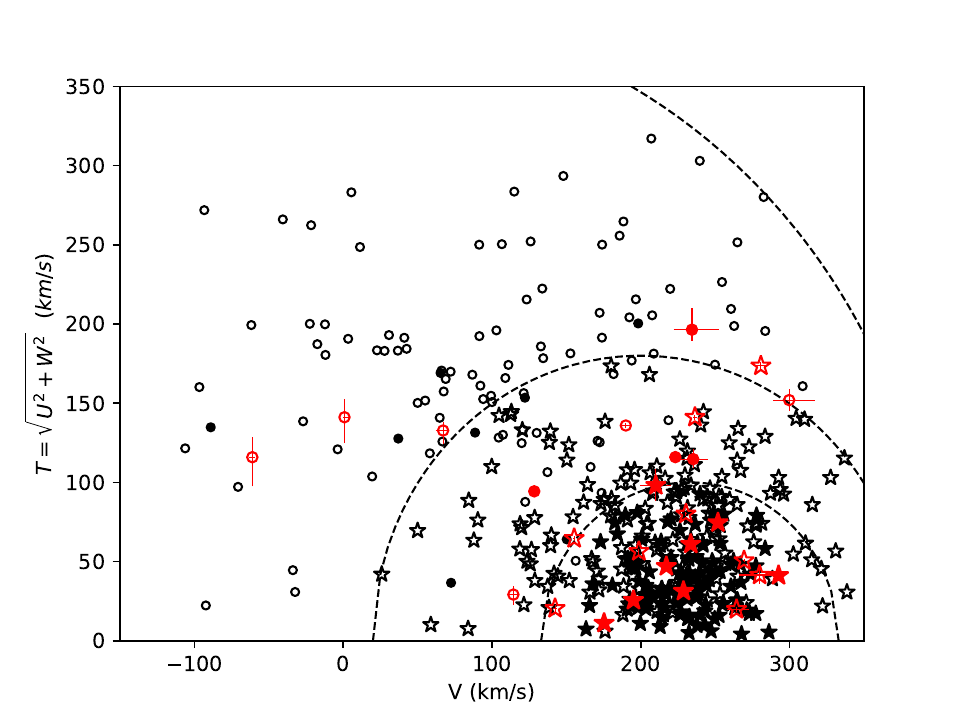}
\caption{The Toomre diagram shows the kinematics of the sample. The symbols have the same meaning as defined in Fig. \ref{RZ}. The dotted lines correspond to predictions of Toomre diagrams from the Besan\c{c}on Galactic models \citep{robin03} for the thin disk ($\leq 100\,\kmsec$), thick disk ($\leq 180\,\kmsec$) and halo stars ($\leq 400\,\kmsec$).}
\label{Toomre}
\end{figure}

\section{Kinematics}
\label{kinematics}

With information about Galactic 3D space velocities and orbits, an attempt may be made to connect both classes and individual stars with parent populations. In the following we describe our classification scheme, and then discuss the classifications in terms of spatial and kinematical distributions. 

\subsection{Classification.}

The population classification described in \cite{martin17a} has been augmented with a classification for the bulge.  We define the extent of the bulge based on estimates for the bulge and bar from \citet{galreview16} who state that it is difficult to separate the outer bulge from the bar; we therefore do not distinguish the bar from the bulge.

Broadly:
\begin{enumerate}
\item Stars with low eccentricity and low inclination orbits are identified with the Galactic disk. 
These are further subdivided into
\subitem \textbf{Thin Disk (TH):} stars with $Z_{\rm max}\, < 1.2$\,kpc and falling within the thin disk regime in the Toomre diagram.
\subitem \textbf{Thick Disk (TK):} stars with $Z_{\rm max}\, > 1.2$\,kpc and falling within the thick disk regime in the Toomre diagram.\\
\item Stars with high eccentricity orbits and/or high orbital inclination are identified with spherical components. They are divided into:
\subitem \textbf{Bulge (B):} stars with low angular momentum, which come in close proximity to the Galactic Centre and stay in the inner disk, i.e. $J_z < 1000$ kpc \kmsec, $|X|\, \leq 4$ kpc, $|Y|\,\leq 1.5$ kpc and $|Z|\,\leq 1.5$ kpc, 
or have $R_{a}<5\,$kpc and $e>0.4$. \\
\subitem \textbf{Halo (H): } stars with $Z_{\rm max}\ >$ 1.2\ kpc and $R_{a} > 7$\,kpc or stars with high orbital inclination.

\end{enumerate}

The classifications derived using the above criteria are shown in Table \ref{A2}. 
A justification for these criteria will be given in the following subsection. 
The orbits for stars which clearly do not match any of the above categories were classified by inspection. 
A similar classification scheme was applied to the stars in the secondary sample.

\subsection{Justification.}

Figure \ref{RZ} shows the Z-distribution of the sample members as a function of galactocentric distance $R$. 
The EHe sample appears to be centred on the disk with a Gaussian distribution along the Z axis, skewed slightly to negative Z.
The Z-distribution of stars in all other classes is broadly similar, although skewed in the opposite sense.  
The scale height of both Gaussian distributions appears to exceed 2 kpc, so the samples as a whole (EHe and all others) cannot belong to the thin disk alone. 

The thin and thick discs kinematically overlap one another in a way that makes it difficult to find a better selection criteria capable of distinguishing them. 
\citet{bovy11} suggests that there may not be a separation between thin and thick discs implying that the transition between them is a continuum.  
Our adopted definition for the  thin disk extent, $|Z_{\rm max}|<1.2$\,kpc, is based on a relationship between the well-established scale heights of the thin disk and the thick disk\, \citep{ma17}, $W$ values, and their number densities\, \citep{bensby14}. \\
We define 
\begin{equation}
    p(z) = n\,e^{-(z/h)^2} 
\end{equation}
as the probability of a star being found at height $z$ in a population with a Gaussian distribution $e^{-(z/h)^2}$, where $h$ is the scale height and $n$ is the normalisation fraction of the population.
Based on this relationship, we can obtain the likelihood that a star at $Z_{\rm max}$ belongs to either the thin disk or the thick disk, such that stars with $|Z_{\rm max}| \, =1.2$\,kpc have an equal likelihood $(\sim 1)$ of being either. 
The $\log$ likelihood of a star belonging to the thick disk rather than the thin disk, $f = p({\rm thick})/p({\rm thin})$, is shown in Table \ref{A2}.
Thus, the thin and thick disk assignments are $>70\%$ secure in all cases except V652\,Her, V2076\,Oph, and BPS\,CS\,22940--0009, which all have a $>33\%$ chance of being mis-classified. 

Figure \ref{Rapzmax} shows the apocentric positions of the sample members in terms of $|Z_{\rm max}|$ and $R_{\rm a}$. 
This representation helps to separate classes because it combines the full orbital information with the spatial information.     
It clearly distinguishes the thin disk EHes from the thick disk and halo, but is not so good at separating out thick disk and halo stars, or thin disk and bulge stars. 
The same holds for stars in the secondary samples. 
There is a deficit of stars with high $\rm Z_{\rm max}$ and low $\rm R_{\rm ap}$. This `zone of avoidance' would correspond to stars in highly inclined orbits which are not expected to be numerous. 

A method frequently used for population classification compares the Z-component of angular momentum ($J_z$) with orbital eccentricity (Fig.\,\ref{eJ}) \citep[cf.][]{pauli03}.
Due to the low eccentricity of the disk stars, they tend to cluster around $e\approx0.2\pm0.1$ and $J_{z}\approx2000\pm500$. 
This diagram was originally devised to identify thick disk stars in the local neighbourhood; because our sample volume is much larger than \citet{pauli03}, some thick disk EHes lie outside  Pauli's region 2.  

Figure \ref{kin} shows the kinetic energy $2 E_{\rm kin}/m = U^{2} + V^{2} +W^{2}$ as a function of  the rotational component V. 
Contours show the velocity perpendicular to Galactic rotation at certain values, where $V_{\bot} =(U^{2} + W^{2} )^{1/2}$. 
Any value for which $V_{\bot} >$ 100 $\rm km\,s^{-1}$ implies a non-disk orbit. 
The higher the value of $2E_{\rm kin}/m$, the more the orbit deviates from a circular orbit. 
This highlights stars which do not belong to the disk. 
We note a number of stars in retrograde motion, particularly the EHe star EC 19529--4430. 

The Toomre diagram (Figure \ref{Toomre}) has been used to classify the stars into their populations using their space velocities $T \equiv V_{\bot} =(U^{2} + W^{2} )^{1/2}$ versus $V$.  Based on the predictions from the Besan\c{c}on Galactic models \citep{robin03}, we identify stars within 100 \kmsec of the solar motion to be likely thin disk stars, within 180 \kmsec to be likely thick disk stars and those within 400 \kmsec to be halo stars.

A useful tool in identifying halo stars with low eccentricities is the `normalized z-extent', which can be substituted as a measure of the orbital inclination, given by
\begin{equation}
    z_{\rm n} = \frac{ Z_{\rm max}}{ R(Z_{\rm max})}
\end{equation}
where $R(Z_{\rm max})$ is the galactocentric distance when the star is at $Z_{\rm max}$.

As this paper was being completed, we received a preprint by  \citet{tisserand23} presenting a similar analysis for HdC, RCB and EHe stars. 
Their sample of 16 EHe stars compares with our sample of 27 EHe stars owing mainly to our inclusion of recent discoveries from SALT \citep{jeffery21a}.  
On comparison of the population classifications for EHes, both papers agree on memberships for the thin disk and bulge stars. 
\citet{tisserand23} did not identify any EHes in the halo, with the possible exception of V2205 Oph. 
We classify V2205 Oph as a halo star (Fig. \ref{galpy}) on the basis of its high eccentricity ($0.57\pm0.07$) and $R_{a}$ ($9.39 \pm 0.53$ kpc).  
3 halo stars in our sample (BD +37\,1977, EC\,19529--4430 and EC\,20236--5703) were not included by \citet{tisserand23}.
4 stars, V2076 Oph, CoD --46\,11775, BD +10\,2179 and CoD --48\,10153, are classified as thick disk by \citet{tisserand23}.  
We classify V2076 Oph as thin disk because of its small $Z_{\rm max}=0.86\pm0.03$\,kpc. 
CoD -46\,11775 lies in the region that we have defined for the bulge (X = $-3.59^{+0.17}_{-0.21}$\,kpc, Y = $-1.32^{+0.05}_{-0.06}$\,kpc, Z = $-0.78^{+0.03}_{-0.03}$\,kpc). 
Although BD +10\,2179 falls within the thick disk region in the Toomre diagram, its high eccentricity ($0.53\pm 0.02$) and $R_{a}$ ($10.59\pm0.15$\,kpc) make it more likely to be a halo star. 
Finally, CoD --48\,10153 falls outside the thick disk regime in the Toomre diagram. 
However, its highly eccentric orbit ($0.51\pm0.02$) and a very low Galactic rotational velocity ($V = 0.9^{+11.8}_{-15.9}\,\kmsec$) clearly make it a halo star.
\begin{table}
\caption[]{Population statistics for various classes of hydrogen-deficient star.}
\label{t:stats}
\centering
\renewcommand{\arraystretch}{1.2}
\begin{tabular}{  lccccc}
\hline
Category&	N&	Thin Disk&	Thick Disk&	Halo &	Bulge\\
\hline
EHe     &27 &8 &9  &6  &4\\[2mm]
RCB    &38 &9 &13  &8  &8\\
HdC     &8 &3 &3 &2 &0\\
He-sdO  &181&41&104 &36&0\\
He-sdOB &135&40 &47 &48 &0\\
He-sdB  &17 &13 &3  &1  &0\\
\hline
\end{tabular}
\end{table}

\subsection{Statistics.}

Table \ref{t:stats} compares the numbers of stars classified as thin or thick disk, halo or bulge stars within various classes of hydrogen-deficient star.  
Five additional subclasses have been classified in exactly the same way as the EHes. 
Of these, the helium-rich subdwarfs (He-sdB, He-sdOB, He-sdO) are intrinsically fainter than the other classes. The absence  of bulge stars in these classes reflects the fact that they are not represented in our sample at such distances. 
O(He) and PG1159 subclasses were also considered, but are too few to provide useful samples.  

It is highly significant that the EHe sample includes stars from the thin disk, the thick disk, the halo and the bulge. 
We shall return to this in \S\,5.  

Given the sample sizes, the distributions of RCB, HdC and He-sdO stars amongst the four components are not significantly different from that of the EHes. 
They are dominated by thick disk stars, with thin disk, halo and bulge stars roughly in the ratio 23:53:20:4. 
The HdC sample shows the largest departure from this ratio, but is roughly consistent given the sample size.  
The He-sdO sample is 7 times larger than the EHe sample. 
Thus, the hypothesis that EHe stars may be connected to 
the RCB, HdC and He-sdO classes cannot be excluded. 

The He-sdB sample is heavily dominated by thin disk stars, but the sample size is small.  
The He-sdOB sample is balanced across thin disk, thick disk and halo and is 4 times larger than the EHe sample. 
It therefore appears that the He-sdB and He-sdOB samples do not share the characteristics of the EHe and He-sdO samples. 
The helium-rich subdwarf samples can be further subdivided into, {\it inter alia}, intermediate helium-rich and extreme-helium rich subclasses \citep{jeffery21a}. 
Detailed analysis of these classes will therefore be deferred.

\section{Luminosities and radii}
\label{luminosities&radii}
One method to derive the luminosity is to obtain the absolute visual magnitude (S. \ref{Gaia_colors}), and then to obtain the absolute bolometric magnitude by applying a bolometric correction (BC). 
Since commonly used values for the BC \citep{BC1967,lang74,Martins05,Nieva13} are derived from hydrogen-rich stars, these might not be applicable to hydrogen-deficient stars. 
An alternative method is to deduce the angular diameter of the star ($\theta$) from the spectral energy distribution (SED), and thence to obtain the actual diameter by multiplying by the distance. 
If the effective temperature is known, the luminosity follows. 
In order to apply this method, a theoretical energy distribution from a model atmosphere is convolved with an extinction function and scaled to fit the observed SED. 
Principal unknowns are the effective temperature of the model atmosphere $T_{\rm eff}$, the extinction coefficient $E_{\rm (B-V)}$ and  $\theta$. 
Other parameters which affect the theoretical model include surface gravity, metallicity and microturbulent velocity. 
In general these cannot be determined from the SED and must be selected by some other means.

{ 
\renewcommand{\arraystretch}{1.3}
\begin{table*}
\caption{Atmospheric parameters, extinction, magnitudes, luminosities and radii for Extreme Helium stars, ordered by effective temperature. The sample has been divided into those stars with $L>2500\Lsolar$ above and those with $L<2500\Lsolar$ below. The sources cited for $T_{\rm eff}$ are also sources for  $\log g$ and extinction $E_{\rm (B-V)}$, where available. '*' indicates a missing error for temperature or gravity; we assume $\pm\,2 \,\%$ for temperature and $\pm 0.25$ for $\log g$. \dag indicates an extinction estimated from \citet{schlegel98}. G and V magnitudes are from \citet{gaia23.dr3}. Masses $M_{L}$ and $M_{g}$ obtained from the $M_{c} - L_{s}$ relation for helium-shell burning are shown for the high luminosity stars  (Fig. \ref{fig:evM}). The three low-gravity low-luminosity stars from Fig. \ref{fig:evM} are also shown in parenthesis. } 
\label{t:params}
\setlength{\tabcolsep}{5pt}
\begin{tabular}{L{2.5 cm} r@{$\pm$}l r@{$\pm$}l ccc r@{$\pm$}l r@{$\,$}l cc  }
\hline
Star & \multicolumn{2}{c}{\Teff} &  \multicolumn{2}{c}{$E_{\rm(B-V)}$} & G & V & Luminosity & \multicolumn{2}{c}{$\log g$} & \multicolumn{2}{c}{Radius }& $M_{L}$& $M_{g}$\\
  &  \multicolumn{2}{c}{(K)} &  \multicolumn{2}{c}{}  & (mag) &(mag)& (log L$_{\odot}$) &\multicolumn{2}{c}{(cm s$^{-2}$)} & \multicolumn{2}{c}{(R$_{\odot}$)}& M$_{\odot}$& M$_{\odot}$\\
\hline
NO Ser & 11750 & 250$^{1}$ & 0.498 & 0.010 & 10.11 & 10.25 & 3.60$^{+0.15}_{- 0.01 }$ & 2.30 & 0.40 & 15.30& $^{+2.07}_{-0.50 }$&$0.60^{+0.03}_{-0.01}$&$0.53^{+0.05}_{-0.04}$\\
V2244 Oph & 12750 & 250$^{1}$ & 0.500 & 0.020 & 10.85 & 10.94 & 3.81$^{+0.22}_{-0.02 }$ & 1.75 & 0.25 & 16.70& $^{+3.80}_{-0.91 }$&$0.66^{+0.05}_{-0.01}$&$0.67^{+0.08}_{-0.06}$\\
PV Tel & 13750 & 400$^{3}$ & 0.130 & 0.010 & 9.23 & 9.26 & 3.83$^{+0.17}_{- 0.06 }$ & 1.60 & 0.25 & 14.60& $^{+2.17}_{- 0.19 }$&$0.64^{+0.04}_{-0.01}$&$0.76^{+0.13}_{-0.08}$\\
LSS 99 & 15330 & 500$^{4}$ & 0.877 & 0.038 & 12.05 & 12.27 & 4.03$^{+0.22}_{-0.22 }$ & 1.90 & 0.25 & 14.80& $^{+2.99}_{-0.98 }$&$0.69^{+0.08}_{-0.05}$&$0.71^{+0.10}_{-0.06}$\\
LSS 4357 & 16130 & 500$^{4}$ & 0.628 & 0.040 & 12.43 & 12.54 & 3.84$^{+0.30}_{-0.05 }$ & 2.00 & 0.25 & 10.90& $^{+3.25}_{-1.10 }$&$0.65^{+0.08}_{-0.01}$&$0.70^{+0.10}_{-0.06}$\\
V1920 Cyg & 16300 & 900$^{3}$ & 0.310 & 0.020 & 10.26 & 10.30 & 3.94$^{+0.19}_{- 0.12 }$ & 1.70 & 0.35 & 11.70& $^{+1.38}_{- 0.35 }$&$0.67^{+0.06}_{-0.03}$&$0.89^{+0.36}_{-0.16}$\\
CoD --46 11775 & 18300 & 400$^{3}$ & 0.161 & 0.015 & 11.17 & 11.20 & 3.57$^{+0.15}_{- 0.22 }$ & 2.20 & 0.20 & 6.20& $^{+0.72}_{- 1.28 }$&$0.61^{+0.03}_{-0.03}$&$0.71^{+0.08}_{-0.05}$\\
EC 19529--4430 & 18540 & 90$^{6}$ & 0.050 & $0.012^{\dag}$ & 11.77 & 11.81 & 3.41$^{+0.34}_{-0.14 }$ & 3.42 & 0.04 & 5.00& $^{+2.21}_{-0.86 }$&$0.57^{+0.06}_{-0.02}$&$0.49^{+0.01}_{-0.01}$\\
V2205 Oph & 20277 & 550$^{7}$ & 0.300 & 0.040 & 10.48 & 10.51 & 4.28$^{+0.21}_{- 0.20 }$ & 2.55 & 0.10 & 11.60& $^{+1.85}_{- 2.27 }$&$0.78^{+0.11}_{-0.08}$&$0.66^{+0.02}_{-0.02}$\\
LSS 5121 & 29772 & 1830$^{7}$ & 0.650 & 0.020 & 13.16 & 13.23 & 3.91$^{+0.32}_{- 0.01 }$ & 3.00 & 0.50 & 3.40& $^{+0.93}_{-0.39 }$&$0.66^{+0.10}_{-0.01}$&$0.77^{+0.30}_{-0.14}$\\
V2076 Oph$^{*}$ & 34000 & 680$^{13}$ & 0.450 & 0.020 & 9.79 & 9.82 & 4.20$^{+0.11}_{- 0.04 }$ & 2.80 & 0.25 & 3.65& $^{+0.38}_{- 0.07 }$&$0.75^{+0.04}_{-0.02}$&$1.10^{+0.53}_{-0.25}$\\
BD +37 $442^{*}$ & 48000 & 960$^{15}$ & 0.090 & $0.002^{\dag}$ & 9.94 & 10.01 & 3.72$^{+0.10}_{- 0.07 }$ & 4.00 & 0.25 & 1.06& $^{+0.10}_{- 0.07 }$&$0.62^{+0.02}_{-0.02}$&$0.67^{+0.09}_{-0.06}$\\
BD +37 $1977^{*}$ & 48000 & 960$^{15}$ & 0.030 & $0.004^{\dag}$ & 10.13 & 10.21 & 3.96$^{+0.12}_{- 0.13 }$ & 4.00 & 0.25 & 1.39& $^{+0.16}_{- 0.17 }$&$0.68^{+0.03}_{-0.03}$ &$0.67^{+0.09}_{-0.06}$\\[2mm]
FQ Aqr & 8750 & 250$^{1}$ & 0.100 & 0.020 & 9.47 & 9.51 & 2.93$^{+0.14}_{- 0.14 }$ & 0.75 & 0.25 & 12.90& $^{+1.13}_{- 1.46 }$ & $\left(0.51^{+0.02}_{-0.01}\right)$ & $\left(0.80^{+0.16}_{-0.1}\right)$\\
LS IV --14 109 & 9500 & 250$^{1}$ & 0.340 & 0.050 & 10.99 & 11.08 & 2.61$^{+0.12}_{- 0.13 }$ & 0.90 & 0.20 & 7.50& $^{+0.49}_{- 0.66 }$ & $\left(0.48^{+0.01}_{-0.01}\right)$ & $\left(0.79^{+0.11}_{-0.08}\right)$ \\
CoD --48 10153 & 10600 & 250$^{2}$ & 0.450 & 0.040 & 11.33 & 11.44 & 3.04$^{+0.17}_{- 0.18 }$ & 1.00 & 0.50 & 10.00& $^{+1.35}_{- 1.62 }$ & $\left(0.52^{+0.02}_{-0.02}\right)$ & $\left(0.85^{+0.85}_{-0.21}\right)$ \\
HD 124448 & 15800 & 400$^{3}$ & 0.092 & 0.010 & 9.94 & 9.97 & 2.82$^{+0.08}_{- 0.10 }$ & 2.10 & 0.25 & 3.43& $^{+0.15}_{- 0.22 }$\\
BD +10 2179 & 17300 & 300$^{5}$ & 0.023 & 0.001 & 9.92 & 9.97 & 3.11$^{+0.15}_{- 0.14 }$ & 2.80 & 0.10 & 4.00& $^{+0.55}_{- 0.50 }$\\
V652 Her & 20950 & 70$^{8}$ & 0.060 & 0.010 & 10.51 & 10.56 & 2.78$^{+0.09}_{- 0.06 }$ & 3.46 & 0.05 & 1.88& $^{+0.17}_{- 0.14 }$\\
BX Cir & 23390 & 90$^{9}$ & 0.239 & 0.008 & 12.54 & 12.57 & 3.13$^{+0.15}_{-0.04 }$ & 3.38 & 0.02 & 2.24& $^{+0.40}_{-0.11 }$\\
GALEX J184559.8--413827 & 26170 & 750$^{10}$ & 0.075 & $0.001^{\dag}$ & 14.57 & 14.62 & 2.74$^{+0.18}_{- 0.22 }$ & 4.22 & 0.10 & 1.02& $^{+0.31}_{- 0.08 }$\\
EC 20236--5703 & 26380 & 130$^{6}$ & 0.055 & $0.012^{\dag}$ & 14.75 & 14.80 & 2.26$^{+0.20}_{- 0.06 }$ & 4.14 & 0.05 & 0.66& $^{+0.14}_{- 0.06 }$\\
LS IV+6 2 & 31800 & 800$^{11}$ & 0.210 & 0.020 & 12.12 & 12.16 & 2.85$^{+0.13}_{- 0.05 }$ & 4.05 & 0.10 & 0.87& $^{+0.09}_{-0.05 }$\\
PG 1415+492 & 32200 & 250$^{12}$ & 0.011 & $0.012^{\dag}$ & 14.27 & 14.34 & 2.21$^{+0.08}_{- 0.10 }$ & 4.20 & 0.10 & 0.41& $^{+0.03}_{- 0.04 }$\\
EC 20111--6902 & 34100 & 110$^{6}$ & 0.036 & $0.012^{\dag}$ & 15.84 & 15.87 & 2.28$^{+0.61}_{- 0.96 }$ & 5.68 & 0.04 & 0.39& $^{+0.39}_{- 0.26 }$\\
BPS CS 22940--0009 & 34970 & 370$^{14}$ & 0.052 & $0.012^{\dag}$ & 13.97 & 14.03 & 2.26$^{+0.08}_{- 0.11 }$ & 4.79 & 0.17 & 0.37& $^{+0.02}_{- 0.04 }$\\
GALEX J191049.5--441713 & 39690 & 90$^{6}$ & 0.085 & $0.006^{\dag}$ & 12.93 & 12.98 & 2.58$^{+0.11}_{- 0.04 }$ & 5.47 & 0.03 & 0.42& $^{+0.05}_{- 0.02 }$\\

\hline
\multicolumn{14}{p{17cm}}{1:\citet{pandey01}, 2:\citet{pandey06b}, 3:\citet{pandey06a}, 4:\cite{jeffery98a}, 5:\cite{kupfer17}, 6:\citet{jeffery21a}, 7:\citet{jeffery01c}, 8:\citet{jeffery01b}, 9:\cite{woolf02}, 10:\citet{jeffery17b}, 11:\cite{jeffery98c},  12:\citet{ahmad03a} 13:\citet{rauch96}, 14:\citet{snowdon22}, 15:\citet{jeffery10} }\\
\end{tabular}
\end{table*}
}

\begin{figure}
\includegraphics[width=0.48\textwidth]{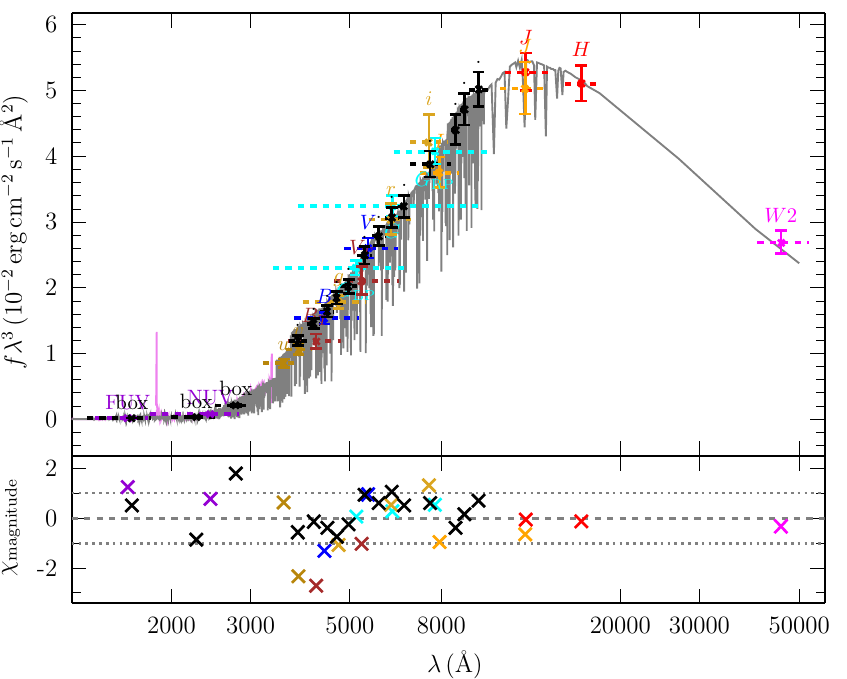}
\caption{ SED fit from the broadband photometry of V2244 Oph with $T_{\rm eff}\,=\,12750\,{\rm K}$ (upper panel). The grey line shows the best-fit model SED. The lower panel shows the residuals. Photometry is taken from: \, \citet{bianchi17}(FUV,NUV),\, \citet{hog00}(B$_{T}$,V$_{T}$),\, \citet{henden15}(B,V),\, \citet{riello21}($\rm G_{BP},\,G,\,G_{RP}$),\, \citet{onken19}(u,v,g,r,i,z),\, \citet{denis05}(I,J), \,\citet{cutri03}(J,H) and \citet{schlafly19}(W2).}
\label{f:SED}
\end{figure}

Photometric observations from a number of sources were used to fit the SED of the EHe stars using the model-atmosphere fitting tool {\sc isis}\footnote{http://www.sternwarte.uni-erlangen.de/gitlab/irrgang/stellar.git} \citep{heber18}. These sources include \citet{hog00},\, \citet{cutri03,cutri21},\, \citet{denis05},\, \citet{mcmahon13},\, \citet{henden15},\, \citet{kilkenny16},\, \citep{bianchi17},\, \citet{schlafly18,schlafly19},\, \citet{onken19},\, \citet{magnier20},\, \citet{drlica21},\, \citet{riello21}. 
The theoretical spectra used for the fits were taken from grids of line-blanketed model atmospheres computed in local thermodynamic equilibrium with {\sc sterne}  \citep{behara06a}. 
The model grid used for each star was chosen according to the measured metallicity and microturbulence, where available, or on the basis of the similarity of the star's optical spectrum  to those of other EHes.  

{\sc isis} allows us to solve for several parameters at once, or to keep some fixed and solve for others. 
For hot stars, $T_{\rm eff}$ and $E_{\rm (B-V)}$ are almost degenerate, so free solutions for both are difficult to control
\citep[cf.][]{jeffery01c}.
For consistency, $T_{\rm eff}$ was fixed to the most recent value obtained from optical spectroscopy wherever possible. 
{\sc isis} then solved for $E_{\rm (B-V)}$ and $\theta$.
Hence, we deduce the radius 
\begin{equation}
    {R_{\star}} = \theta d / 2
\end{equation}
and the luminosity 
\begin{equation}
    \frac{L_{\star}}{\rm L_{\odot}} = \left(\frac{R_{\star}}{\rm R_{\odot}}\right)^{2} \, \left(\frac{T_{\rm eff}}{\rm T_{\rm eff \, \odot}}\right)^{4}.
\end{equation}
These values have been computed for the EHe stars and are given in Table \ref{t:params} along with published values for the effective temperatures ($T_{\rm eff}$), surface gravities ($\log g$) and extinction coefficients ($E_{\rm (B-V)}$).
$T_{\rm eff}$ and $\log g$ have been taken from the most recent spectroscopic analyses of optical spectra wherever possible and the sources are cited in Table \ref{t:params}. 
For measurements that do not have a published error, we have assumed an error of $\pm\,2 \%$ for temperature and $\pm$ 0.25 dex for $\log g$. 
Some measurements cited in Table \ref{t:params}  have very small error; these are taken to be formal errors. 
To partially account for likely larger systematic errors, minimum errors of $\pm \, 1 \%$ and 0.05 dex are adopted for $T_{\rm eff}$ and $\log g$ respectively for propagation. 
Figure\,\ref{f:SED} shows an example of an SED fit for the cool EHe V2244\,Oph. None of the SED fits show any evidence for an infra-red excess that might indicate an unseen cool companion or a dust shell.

Since we have the surface gravity and radius for each star, we may in principle deduce the mass 
 \begin{equation}
      M_{\star} = \frac{\rm g}{\rm G} \, R_{\star}^{2}.  
 \end{equation}
In practice, whilst the errors in radius are usually smaller than $\pm\,10\%$, their contribution doubles in the derivation of mass and the errors in surface gravity are mostly between $\pm\,25\%$ and $\pm\,80\%$. Consequently, the mass errors ($1\sigma$) exceed $100\%$ for half of the EHe sample, and are less than $50\%$ for only 5 EHes.
The deduced masses run from $0.02^{+0.015}_{-0.003} -  2.71^{+3.62}_{-5.44}$ \Msolar\, and  12 cluster between  0.2 and 0.5  \Msolar. 
In view of the errors and the spread, we attach little confidence to these data. 
Likely reasons include the difficulty of measuring surface gravities and the presence of large amplitude pulsations in these  luminous stars; unless angular radii and surface gravities are obtained at equivalent phases in approximately hydrostatic equilibrium, significant systematic errors will occur. 
The issues are examined in detail for the case of V652\,Her by \citet{jeffery22a}.

\begin{figure}
\includegraphics[width=0.53\textwidth]{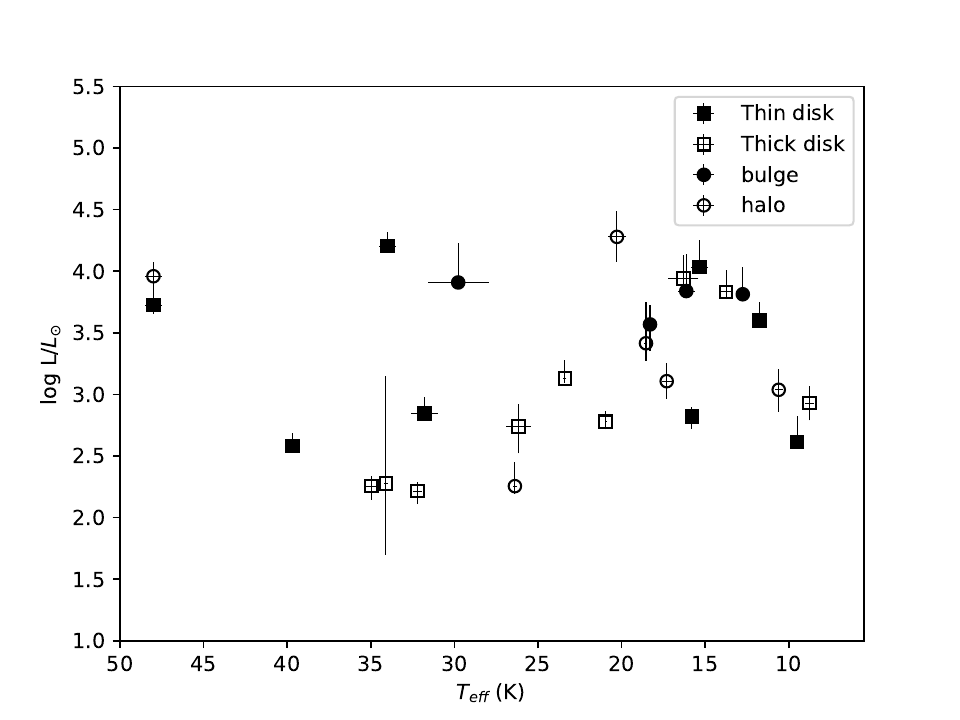}
\caption{Luminosities of EHes derived from {\it Gaia} distances and SED-based angular diameters as a function of effective temperature. Symbols represent the population assignments given in Table~\ref{t:params} (see key).}
\label{fig:LT}
\end{figure}


\begin{figure}
\includegraphics[width=0.53\textwidth]{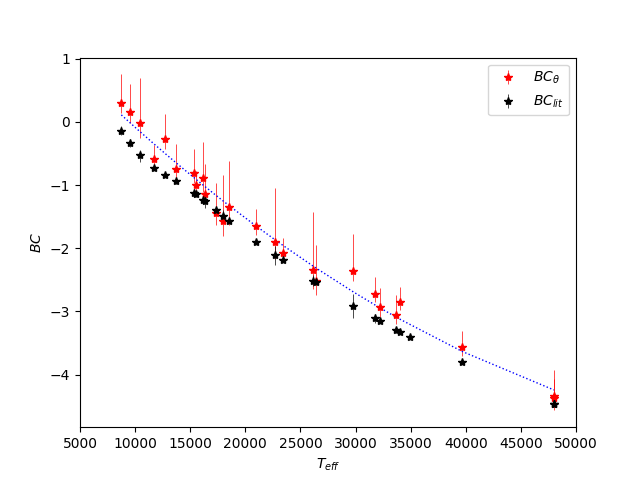}
\caption{Comparison between BC adopted from \citet{BC1967},\,\citet{lang74},\,\citet{Martins05} and \citet{Nieva13} ($\rm BC_{lit}$) and BC obtained using the spectral energy distributions (${\rm BC}_{\theta}$) of the EHe stars in the temperature range 8750\,K - 48000\,K. The blue dotted line depicts a low order polynomial fit to the data.}
\label{fig:TBC}
\end{figure}

Figure \ref{fig:LT} shows the luminosities as a function of  effective temperature. This suggests that EHes may be divided into two luminosity groups, with a dividing line at $L\approx2\,500\Lsolar$. 
The more luminous group $L\approx2\,500 - 20\,000\Lsolar$ shows roughly similar luminosity over the entire temperature range.
The less luminous group $L\approx100 - 2\,500\Lsolar$ shows a progression towards lower luminosity with increasing temperature.
The two groups are also identified in Table \ref{t:params}.  
It is significant that both groups include stars from the thin disk, the thick disk, and the halo.  

\subsection{Bolometric correction for EHe stars}

To determine whether the bolometric corrections available in literature for stars of these temperature ranges could be applicable to EHes, we converted the SED luminosities to $M_{\rm bol \star}$ using
\begin{equation}
    M_{\rm bol \star} = -2.5 \log_{10} \left(\frac{L_{\star}}{\rm L_{\odot}} \right) + {\rm M_{\rm bol \odot}}
\end{equation}
and then compared with the absolute visual magnitude from Eq. \ref{eq:mv} to obtain the bolometric correction 
\begin{equation}
    \rm BC = M_{\rm bol \star} - M_{V}.
\end{equation}

Figure \ref{fig:TBC} compares the EHe bolometric corrections with the bolometric corrections from \citet{BC1967,lang74,Martins05,Nieva13}. 
The correction for helium stars seems to be more positive than the literature values by about 0.3 mag. 
This is probably due to the absence of Balmer line and continuum absorption in the EHe stars. 
The BC for EHes can be expressed as a function of $T_{\rm eff}$ by the polynomial fit shown in Fig.\,\ref{fig:TBC}: 
\begin{equation}
    BC = 1.59  -1.79\,(T_{\rm eff}/10^4) + 0.122\,(T_{\rm eff}/10^4)^2. 
\end{equation} 
This correction can also be used to obtain luminosities for stars with insufficient photometric data for an SED fit.




\begin{figure*}
\includegraphics[width=0.49\textwidth]{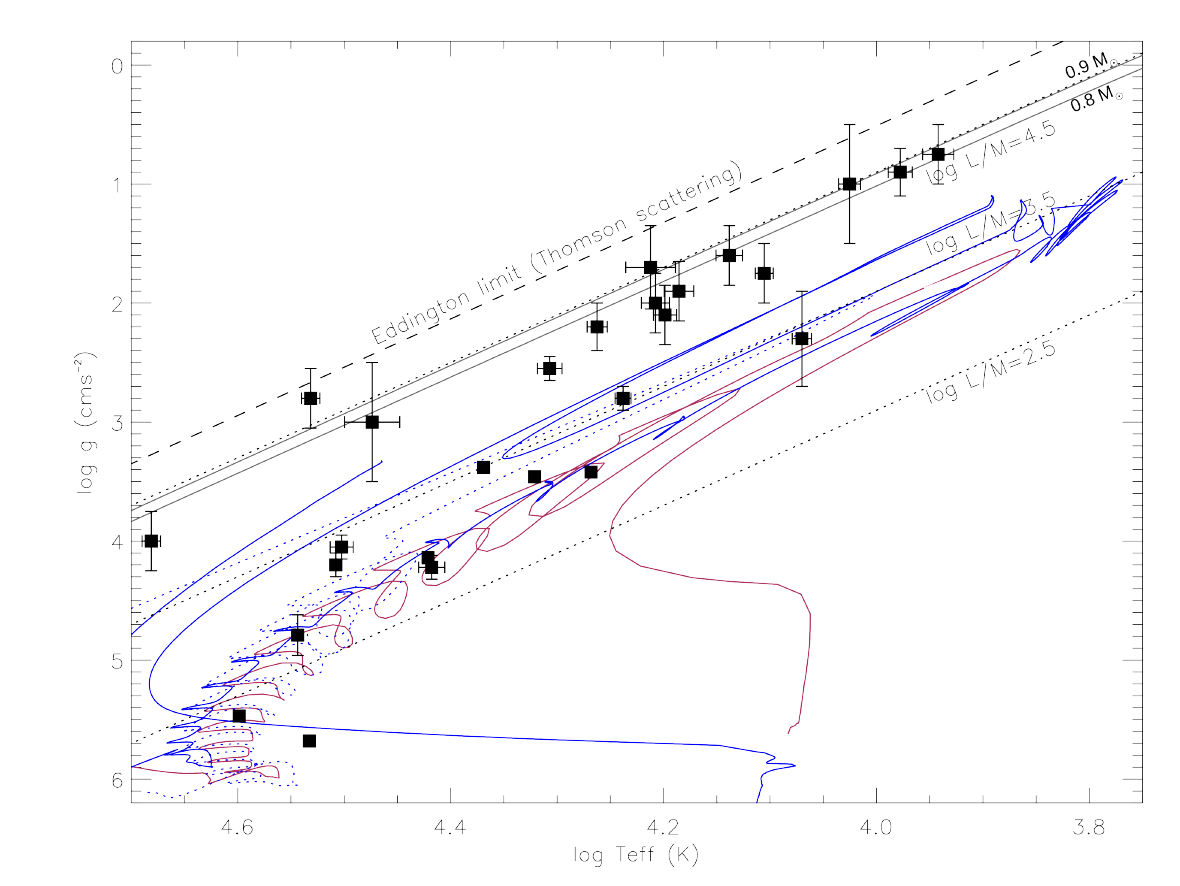}
\includegraphics[width=0.49\textwidth]{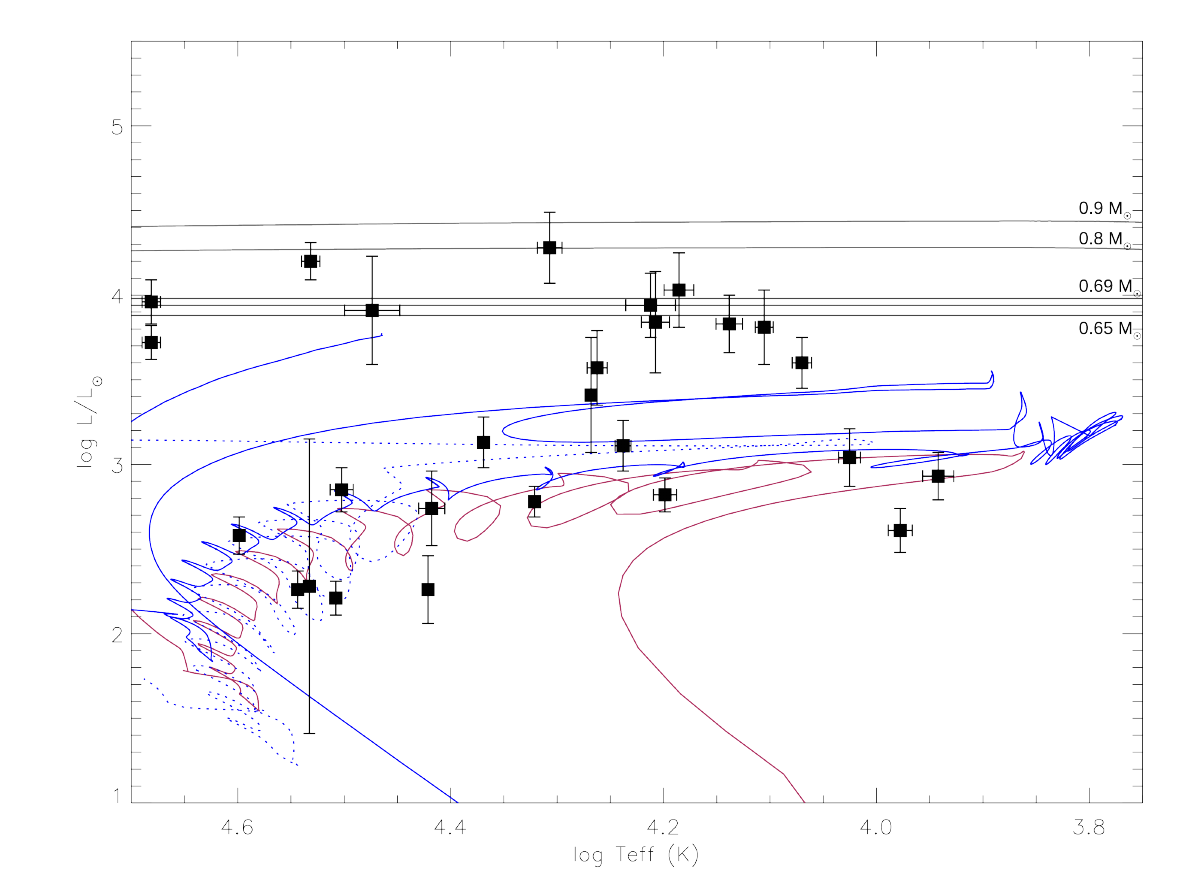}
\caption{A comparison of the post-merger evolution of double white dwarfs with the observed properties of Extreme Helium stars. 
Left: the Kiel ($g-T_{\rm eff}$) diagram. 
The black dotted lines indicate equal L/M ratio as marked. 
The grey lines show the post-merger evolution tracks for models of CO+He WD mergers from \citet{saio02} with product masses $0.6+0.2$ and $0.6+0.3$ $\rm M_{\odot}$. 
The blue lines show tracks for He+He WD mergers from \citet{saio00} with the dotted line corresponding to $\rm 0.3 + 0.2\,M_{\odot}$ and solid corresponding to $\rm 0.4 + 0.3\,M_{\odot}$. 
The red line shows the track for a $0.3 + 0.25\, \rm M_{\odot} $ He+He WD merger from \citet{zhang12a}
Right: The Hertszprung-Russell ($L-T_{\rm eff}$) diagram, 
plotted with the same evolution tracks. 
Dark grey lines show CO+He WD post-merger tracks from \citet{schwab19} with final masses 0.69, 0.67 and 0.65 $\rm M_{\odot}$.}
\label{fig:evLgT}
\end{figure*}


\section{Inferences on Evolution, Age, and Numbers}

With spatial and kinematic data, as well as luminosities, we are in a position to discuss the age, origin and evolution of EHes. 

Using their distribution and radial velocity in the Galactic frame, \citet{drilling86} argued that EHe stars belong to the Galactic bulge.  
\citet{iben85} derived a scale height 1700\,pc for the EHe star population, which would be commensurate with a thick disk population. 
The current analysis shows that EHes include stars representative of all the major Galactic populations, including the thin and thick disk as well as the bulge and halo. 
Thus EHes must arise from star formation at many epochs of Galactic evolution. 
The {\it Gaia} analysis also shows that EHes form two luminosity groups, implying more than one formation channel. 

\subsection{Evolution}

It has been proposed that RCB and EHe stars form as a result of the merger of two white dwarfs (WDs) \citep{webbink84,saio02}. 
Post-merger evolution depends on the nature of the WDs involved; for example:
\begin{itemize}
    \item He+He WD$\xrightarrow{}$ ?? $\xrightarrow{}$ EHe $\xrightarrow{}$ He-sdO $\xrightarrow{}$ DO WD
    \citep{saio00,zhang12a}\footnote{\citet{iben90} proposed that a He+He WD merger would produce an sdB star; in fact, Iben's model cannot account for the classical hydrogen-rich sdB stars but is equivalent to this sequence.}
    \item He+He WD$\xrightarrow{}$ ?? $\xrightarrow{}$ EHe $\xrightarrow{}$ He-sdO $\xrightarrow{}$ RCB $\xrightarrow{}$EHe $\xrightarrow{}$ He-sdO$^{+}$ $\xrightarrow{}$ O(He) $\xrightarrow{}$ DO WD
    \citep{zhang12b}
    \item Co+He WD$\xrightarrow{}$ RCB $\xrightarrow{}$ EHe $\xrightarrow{}$ He-sdO$^{+}$ $\xrightarrow{}$ O(He) $\xrightarrow{}$ DO WD
    \citep{webbink84,saio02,zhang14,schwab19}
\end{itemize}
The middle sequence may be impossible to distinguish from either the first or last sequence, having the characteristics of both. 
It is not clear what the immediate product of a He+He WD merger will be; hence '??' above.
Its lifetime will be short, being only the lifetime of a helium shell-flash cycle at the start of core helium burning. 
It may well be a cool EHe or HdC star.
In contrast, the lifetime of the immediate He+CO WD merger product is relatively long, being the lifetime for the helium-burning shell to consume available fuel above the carbon-oxygen core; this may be some 0.3 - 0.4 \Msolar. 
It is reasonable that these may be RCB or HdC stars. 

Given the two EHe luminosity groups, it seems likely that the EHe sample includes post-merger products from both CO+He WD binaries {\it and} He+He WD binaries.
The lower luminosity EHes ($L>2\,500\Lsolar$) and He-sdO's have already been identified with the He+He WD merger sequence \citep{saio00,zhang12a}. 
The higher luminosity EHes ($L>2\,500\Lsolar$) are identified with the merger of a CO+He WD binary \citep{saio02,zhang14,schwab19}, evolving via the low gravity hot sdO domain (He-sdO$^{+}$ and O(He) stars) to become hot WDs.
\citet{clayton07} established that the cooler RCBs lie on the same pathway. 
A strict boundary cannot be drawn because it is possible to form He+He systems which are more massive than the lightest CO+He systems, and because the former can emulate the evolution of the latter after completion of core helium burning \citep{zhang12b}.

Figure~\ref{fig:evLgT} compares post-merger  tracks for both He+He and CO+He WD mergers with the {\it Gaia}-derived luminosities as well as the spectroscopic surface gravities of the EHes.  

\begin{figure}
\includegraphics[width=0.48\textwidth]{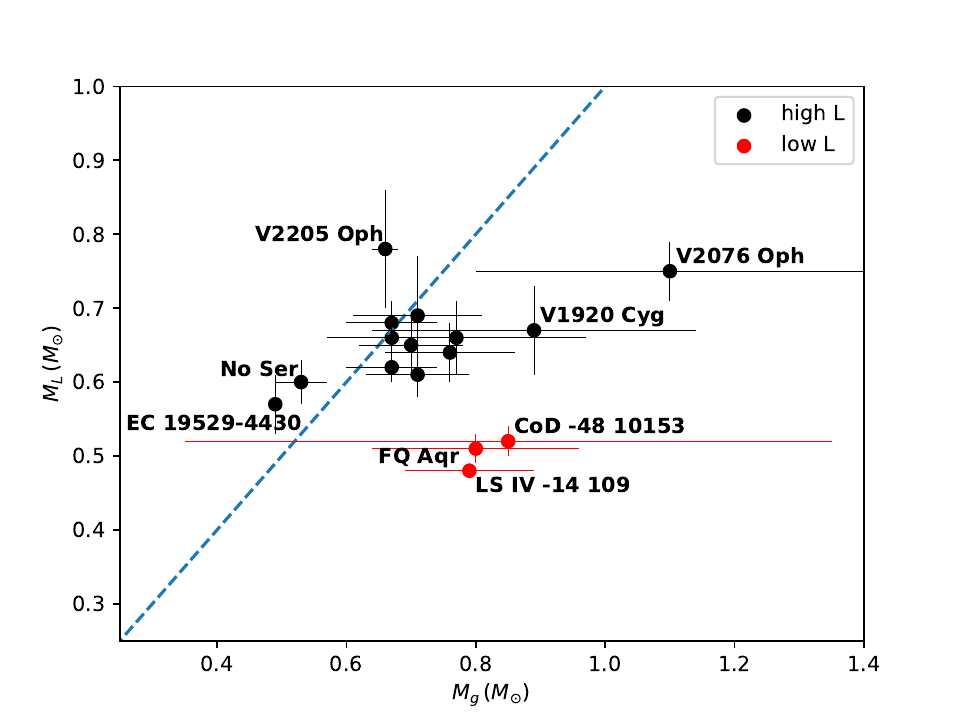}
\caption{EHe masses derived from observed luminosities ($M_L/\Msolar$) and surface gravities ($M_g/\Msolar$) and the $M_c-L_s$ relation appropriate for CO+He WD mergers \citep{jeffery88}.
}
\label{fig:evM}
\end{figure}


The intrinsically brighter EHes have luminosities consistent with post-merger tracks for CO+He WDs in the 0.65 -- 0.9\Msolar\ range, as identified by \citet{saio02,zhang14,schwab19}. 
For these, evolution proceeds via contraction at constant luminosity, approximately on a thermal timescale.  The luminosity reflects the stars' masses, as indicated by the core-mass shell-luminosity ($M_c-L_s$) relation for helium-shell burning stars \citep{jeffery88} which we have verified against the \citet{saio02} and \citet{schwab19} evolution tracks. 

Systems with $L<2\,500\Lsolar$ are more likely to have been produced in He+He systems with total masses $\leq0.7\Msolar$.
Whilst initially bright after shell-helium ignition, these stars contract and dim as the helium-burning shell burns inwards towards the core, eventually reaching the helium main-sequence. 

Estimates for the mass of each EHe star can be obtained from theoretical evolution tracks (Fig.\,\ref{fig:evLgT}) or from the \citet{jeffery88} $M_c-L_s$ relation by separately comparing with the observed surface gravity to give mass $M_g$, and luminosity to give $M_L$.  
 
Values of $M_L$ and $M_g$ obtained from the $M_c-L_s$ relation for high-$L$ EHes ($L>2\,500\Lsolar$) are shown in Table\,\ref{t:params} and compared in Fig.\,\ref{fig:evM}. 
V2244\,Oph, LSS\,99, LSS\,4357, BD+37\,442 and BD+37\,1977 have $M_g$ and $M_L$ in good agreement between 0.62 and 0.71 \Msolar.  
LSS\,5121, CoD\,-46\,11775 and PV\,Tel have $M_L$ in the range $0.59 - 0.66 \Msolar$, but $M_g$ in the range $0.71 - 0.76 \Msolar$.  
NO\,Ser and EC\,19529--4430 have $M_L$ in the range $0.57 - 0.60 \Msolar$, but markedly lower $M_g$.  
For V2205\,Oph, $M_g$ is also too low.
V1920\,Cyg and V2076\,Oph have $M_g$ substantially higher than the corresponding $M_L$.  
Overall, the {\it Gaia}+SED luminosities support masses for post-merger He+CO white dwarfs in the range $0.57-0.78\Msolar$. Significant differences between $M_L$ and $M_g$ should be examined in more detail, particularly with respect to surface gravity and extinction.  

FQ\,Aqr, LS\,IV-14\,109 and CoD\,-48\,10153 are included in Fig.\,\ref{fig:evM} because their surface gravities suggest a high luminosity, which is not supported by the {\it Gaia}+SED measurements. 
The low-$L$ EHes ($L<2\,500\Lsolar$) lie well below the luminosities predicted for CO+He mergers. 
Masses cannot be derived from the post-merger evolution tracks for He+He WDs since, for a given class of model, these are degenerate in $g$ (and $L$) during contraction toward the helium main-sequence \citep{zhang12a}. Moreover,  loops associated with helium-shell flashes broaden the tracks by more than any difference caused by a change in mass. 

According to \citet{tisserand20}, the cool group of RCB stars represents about 2/3 the entire RCB sample, but is under-represented in atmospheric analyses. 
With 118 Galactic RCB stars now known \citep{tisserand20}, substantial numbers of accurate $\teff,g$ measurements for stars within 10\,kpc could transform Fig.\,\ref{fig:evLgT} into a tool that will places useful constraints on evolution models and stellar physics. 
Additionally, with precise bolometric corrections for a larger temperature range, Fig.\,\ref{fig:evLgT} would be expanded to include other sets of helium stars, thereby shedding more light on evolutionary pathways.

\subsection{Age}

The defining timescales for double WD mergers are those of the progenitor binary followed by the orbital-decay time of the double WD system. 
Post-merger evolution timescales are comparatively short, so all current EHes probably formed within the last 10$^4$ -- 10$^6$ y. 
From a composite model for the star-formation history of the Galaxy, \citet{yu11} argue that the majority ($>95\%$) of CO+He double WDs currently alive in the Galactic disc were formed between 4 and 10 Gyr before the present epoch. 
However, the fraction of these that contribute to the present-day CO+He WD merger rate is dominated by star formation between 1 and 3 Gyr before the present. 
Supporting this, \citet{zhang14} show that, for a single starburst, CO+He WD mergers peak some 1\,Gyr after the starburst, declining exponentially by a factor 10 over 3\,Gyr, but then remaining steady up to at least 10 Gyr after the starburst (i.e. at $\approx$1/10 the rate for $\approx10\times$ the time).  
Thus the majority of EHes should come from relatively recent star formation. 
This is supported by the fact that 2/3 of the sample belong to the thin (8) and thick (9) disks (Table\,\ref{t:stats}).
However, the tail of the distribution, which formed up to 10 Gyr  or more in the past, will continue to provide a significant number of recent CO+He WD mergers in all  Galactic populations.  
 
\begin{table}
\caption{Relative frequency of double white dwarf binary mergers by Galactic population and binary white dwarf class at the current epoch based on \citet{yu10,yu11} and normalised to 100 yr$^{-1}$. 
The associated total rate for the Galaxy over all merger types is $\sim 2.8\times10^{-3}$ yr$^{-1}$. }
\label{tab:bsps}
\begin{tabular}{lrrrrr}
\hline
White Dwarfs & Galaxy & Thin Disk & Thick Disk & Halo & Bulge \\
\hline
He+He   & 34  & 17 &  1 & 16 &  8 \\
CO+He   & 14  &  7 &  1 &  5 &  2 \\
CO+CO   & 46  & 22 &  2	& 10 &  5 \\
ONeMG+X &  6  &  3 &  0	&  1 &  1 \\
All     & 100 & 49 &  4	& 32 & 15 \\
\hline
\end{tabular}
\end{table}

\subsection{Numbers}
Predictions for relative numbers of double white dwarf mergers in different populations have been derived in many studies using the methods of binary star population synthesis (BSPS) \citep[{e.g.}][]{yu10,yu11,zhang14}. 
Table\,\ref{tab:bsps} shows the relative merger rates obtained by combining the rates per Galactic component \citep[Table 2]{yu10} with the rates per merger type \citep[Tables 2,3]{yu11}, assuming a quasi-exponential star-formation history for the thin and thick disks, and an instantaneous star-formation history for the bulge and halo.  
These statistics include both stable and unstable mergers \citep[Figs.\,1,24]{zhang14}. 
Only a subset of the latter will give rise to EHe stars, a supernova explosion being an alternative for the most massive mergers.    
There are significant differences between merger-rate predictions. 
\citet[Fig.\,11]{yu11} and \citet[Fig.\,23]{zhang14} show the same quantities, with qualitatively similar, but quantitatively different, results. 
This is not the place to explore the particulars of different BSPS implementations. 
Rather, the intent of Table\,\ref{tab:bsps} is to compare the contributions of He+He and CO+He merger channels and their parent populations with observed EHe numbers (Table\,\ref{t:stats} and Figs.\,\ref{fig:LT},\ref{fig:evLgT}). 

Ideally, merger rates should be combined with the lifetimes of post-merger products to obtain number densities, and adjusted for completeness in order to compare with observed numbers. 
However, even without adjustment, Table \,\ref{tab:bsps} clearly shows a shortage of predicted WD mergers in the thick disk relative to other locations when compared with the observed numbers of EHes, RCBs and He-sdOs (Table\,\ref{t:stats}). 
\citet[Table 1]{yu10} used a model in which the thick disk has 5\% the mass of the thin disk, whilst \citet[Table A.1]{bensby14} cite a ratio closer to 11\%, but this cannot account for all of the shortage in predicted numbers. 
The observed numbers suggest the delay-time distribution for He+He and CO+He WD mergers is too low in the 3 - 10 Gyr range, and hence that the distribution of initial periods of post-common-envelope double WDs is too heavily weighted to short periods. 
Adjusting the common-envelope ejection efficiency in binary star evolution models to produce more widely separated post-common-envelope systems could ameliorate the problem. 
Another explanation could be that, since EHes are rare but luminous, they are seen at greater distances in the thick disk  than in the more heavily obscured thin disk; is there a missing population of EHes close to the Galactic plane?  
We note that the kick velocity imparted to the survivor of a double white dwarf merger due to the destruction of one component is only a few km\,s$^{-1}$ \citep{dan14} and will have little overall effect on the population assignments. 

It has been proposed that EHes could alternatively follow canonical stellar evolution as stripped post-AGB stars or as the remnant of a late thermal pulse in a post-AGB star \citep{iben83}. 
Whilst neither proposal explains the absence of binaries and the surface chemistries of EHes, it is useful to compare with the Galactic distribution of other post-AGB stars. 
A well-defined sample is provided by the planetary nebulae (PNe).
PNe with [WC]-type central stars could represent a sequence of late thermal pulse stars \citep{herwig01,blocker01}. 
PNe in the galactic halo are extremely rare \citep[][3 field PNe]{clegg87}. 
\citet{gorny04} divided the Strasbourg-ESO catalogue of Galactic PNe \citep{acker92} into 263 bulge objects and 878 disk objects. The fraction of [WC]-type PN is similar across both populations at 55/263 for the bulge and 154/878 for the remainder. However, considering selection effects, \citet{gorny04} concluded that the fraction of [WR] PNe is significantly larger in the bulge than in the disk. The San Pedro Mártir kinematic catalogue \citep{lopez12} divided the Galactic PNe sample into 405 thin disk, 24 thick disk, 17 halo and 155 bulge PNes. 
\citet{ali13} found roughly equal numbers of PNe in the thin and thick disks. 
Crudely, the available data suggest a distribution of $\approx$
48:27:1:24 for the thin disk, thick disk, halo and bulge PNe. 
Similarly, the Galactic population of AGB stars is dominated by the thin disk and bulge, with a survey of IRAS-selected AGB stars finding only 15/6039 stars having $|Z|>2.5$\,kpc \citep{jackson02}.
The larger fraction of halo EHe and other hydrogen-deficient stars (Table \ref{t:stats}) contrasts starkly with the deficit of halo PNe and AGB stars. 
It would appear that halo stars massive enough to produce PNe and, by inference, a late thermal pulse, have long since completed their evolution. 
In contrast, double white dwarfs in the halo can continue to produce mergers on a timescales commensurate with the age of the Galaxy. 

\section{Conclusions}

{\it Gaia} DR3 has provided parallaxes and proper motions for  hundreds of millions of stars, including extremely rare hydrogen-deficient stars such as R\,CrB variables, hydrogen-deficient carbn giants, extreme helium stars (EHes), and helium-rich hot subdwarfs.   

In conjunction with the ongoing SALT survey of hydrogen-deficient stars \citep{jeffery21a} we have updated the list of known EHes;
the focus of this paper has been to examine their Galactic distribution and kinematics, and hence learn something about their origin, age, and relationship with other classes of hydrogen-deficient star. 

The {\it Gaia} DR3 data have been processed with \verb|galpy| to obtain complete orbital and spatial information for 27 EHes.
We have also analysed comparison samples of 38 RCBs, 8 HdCs, 181 He-sdOs, 135 He-sdOBs and 17 He-sdBs. 
These data have been used to classify the parent populations for each star. 
EHes can clearly be found in all of the thin disk, the thick disk, the halo and the bulge of the Galaxy.
The same is true, in similar proportion, for RCBs, HdCs and He-sdOs, but the largest fraction of all four classes are found in the thick disk. 
EHes, RCBs and HdCs are luminous and hence have short lifetimes. Their presence in all four components of the Galaxy implies that their progenitor lifetimes must cover a wide range from a few to more than 10 Gyr.  
 
We have combined the {\it Gaia} DR3 distances with analyses of the EHe spectral energy distributions to obtain radii and luminosities.
From the luminosities, we find that the EHes can be divided into two luminosity groups, with the division at $\log L/{\rm L_{\odot}} \approx 3.4$. 
Thin and thick disk and halo populations are all represented in both luminosity groups. 
The existence and population characteristics of both groups are best understood if: \\
a) EHes with lower luminosities ($L<2\,500\Lsolar$) correspond to the post-merger evolution of double white dwarf binaries containing two helium white dwarfs; these are contracting towards the helium main-sequence where they will be observed as He-sdOs, and 
b) EHes with higher luminosities ($L>2\,500\Lsolar$) correspond to the post-merger evolution of double white dwarf binaries containing a carbon-oxygen and a helium white dwarf; these are contracting at roughly constant luminosity to eventually become hot white dwarfs.
We have also derived an equation for the bolometric correction appropriate to hydrogen-deficient stars in the effective temperature range $7\,000 - 50\,000$\,K.  
 
We have briefly discussed the parent populations for both classes of double white dwarf merger. 
Binary star population synthesis predicts that double white dwarfs capable of merging at the current epoch should exist in all Galactic populations. 
However, the predictions favour mergers arising from recent star formation (i.e. thin disk), whereas the statistics favour an older epoch (i.e. thick disk).  
Further analyses of the He-sdO, RCB and HdC populations will be instructive. 

\section*{Acknowledgments}
APM and PM acknowledge studentships from the Armagh Observatory and Planetarium which, in turn, is supported by a core grant from the Northern Ireland Department for Communities. 
CSJ acknowledges support from the UK Science and Technology Facilities Council via UKRI Grant No. ST/V000438/1. 

This work has made use of data from the European Space Agency (ESA) mission {\it Gaia}\footnote{\tt www.cosmos.esa.int/gaia}, processed by the {\it Gaia} Data Processing and Analysis Consortium (DPAC)\footnote{\tt www.cosmos.esa.int/web/gaia/dpac/consortium}. 
Funding for the DPAC has been provided by national institutions, in particular the institutions participating in the {\it Gaia} Multilateral Agreement.

We thank the referee for very helpful comments.


\section*{Data Availability Statement}
The astrometric data underlying this article are available in the {\it Gaia} Archive at ESA and can be accessed via the archive website at {\tt gea.esac.esa.int/archive/}. 
 All other data are available in the publications cited. 

\bibliographystyle{mnras}
\bibliography{ehe} 

\bsp	
\appendix
\section{Secondary Sample}
\label{a:sample}

The stars in the secondary sample defined in Section 2, numbering 383, include members of various classes of hydrogen-deficient stars and are shown in Table\,\ref{T:secondary_detailed}. The method used to deduce population membership is the same as that for EHes (Section 4).  

\begin{table*}
\caption[]{Orbital parameters and Galactic velocities of the secondary sample grouped by class and sorted as in Table \ref{t:gaia}. The population membership has the same meaning as in Table \ref{A2}.}
\label{T:secondary_detailed}

\setlength{\tabcolsep}{5pt}
\tabulinesep=1.5mm
\begin{tabular}{ L{4cm} rrrrr rrrcc}
\hline
Star & $U$ & $V$ & $W$ & $J_z$ & $e$ & R$_{a}$ & R$_p$ & Z$_{\rm max}$ &pop  &class \\
  & \kmsec & \kmsec & \kmsec & kpc \kmsec & & kpc & kpc & kpc &  &\\
\hline
V* XX Cam & $33.1$ & $238.3$ & $3.9$ & $2227.4$ & $0.10$ & $10.77$ & $8.82$ & $0.07$ & TH & RCB\\
V* UX Ant & $-257.9$ & $-93.4$ & $85.8$ & $3964.7$ & $0.11$ & $22.98$ & $18.41$ & $7.97$ & H & RCB\\
V* UW Cen & $-255.2$ & $185.8$ & $16.2$ & $1610.1$ & $0.08$ & $7.68$ & $6.49$ & $1.16$ & H & RCB\\
V* Y Mus & $-215.5$ & $123.3$ & $-1.5$ & $1155.9$ & $0.29$ & $6.93$ & $3.80$ & $0.78$ & H & RCB\\
V* V854 Cen & $-10.9$ & $231.3$ & $-119.3$ & $1462.0$ & $0.43$ & $11.53$ & $4.61$ & $3.18$ & TK & RCB\\
V* Z UMi & $81.5$ & $173.3$ & $29.0$ & $812.6$ & $0.77$ & $15.25$ & $1.96$ & $4.51$ & TK & RCB\\
V* S Aps & $-117.0$ & $242.1$ & $85.0$ & $1524.9$ & $0.10$ & $7.51$ & $6.19$ & $1.54$ & TK & RCB\\
V* R CrB & $-50.3$ & $241.7$ & $34.9$ & $1839.0$ & $0.13$ & $9.12$ & $7.09$ & $0.93$ & TH & RCB\\
V* IO Nor & $-87.6$ & $175.4$ & $16.7$ & $891.5$ & $0.36$ & $5.84$ & $2.75$ & $0.06$ & TK & RCB\\
V* RT Nor & $-78.0$ & $154.5$ & $-6.6$ & $808.1$ & $0.23$ & $4.70$ & $2.92$ & $0.76$ & TK & RCB\\
IRAS 16571-5011 & $-105.0$ & $67.5$ & $-117.3$ & $452.3$ & $0.48$ & $5.15$ & $1.79$ & $2.84$ & H & RCB\\
V* GV Oph & $49.2$ & $150.2$ & $-41.1$ & $445.3$ & $0.46$ & $4.08$ & $1.52$ & $0.98$ & B & RCB\\
SV* HV  7863 & $-101.4$ & $226.2$ & $77.0$ & $889.6$ & $0.19$ & $4.84$ & $3.29$ & $0.73$ & TK & RCB\\
ASAS J171710-2043.3 & $-139.2$ & $65.4$ & $-96.1$ & $23.6$ & $0.91$ & $2.38$ & $0.11$ & $1.86$ & B & RCB\\
V* V2552 Oph & $-21.0$ & $122.1$ & $-152.0$ & $543.7$ & $0.40$ & $4.75$ & $2.03$ & $1.76$ & B & RCB\\
Terz V 2637 & $-112.4$ & $37.0$ & $60.5$ & $-56.3$ & $0.49$ & $1.12$ & $0.39$ & $0.42$ & B & RCB\\
V* V653 Sco & $-246.9$ & $-40.6$ & $99.1$ & $233.1$ & $0.92$ & $13.08$ & $0.51$ & $2.55$ & H & RCB\\
V* WX CrA & $35.0$ & $88.6$ & $-126.7$ & $210.3$ & $0.46$ & $2.64$ & $0.97$ & $1.46$ & B & RCB\\
V* VZ Sgr & $459.0$ & $159.2$ & $175.7$ & $171.9$ & $0.78$ & $4.64$ & $0.58$ & $2.46$ & B & RCB\\
V* RS Tel & $-35.3$ & $72.5$ & $9.7$ & $194.7$ & $0.60$ & $3.60$ & $0.91$ & $1.41$ & B & RCB\\
V* FH Sct & $14.9$ & $267.1$ & $-106.5$ & $876.5$ & $0.12$ & $4.66$ & $3.68$ & $1.34$ & TK & RCB\\
V* V CrA & $-5.1$ & $338.5$ & $30.3$ & $3660.8$ & $0.53$ & $35.04$ & $10.85$ & $3.03$ & TK & RCB\\
V* SV Sge & $25.6$ & $239.9$ & $34.1$ & $1534.7$ & $0.13$ & $7.47$ & $5.73$ & $0.44$ & TH & RCB\\
V* V1157 Sgr & $-1.7$ & $260.5$ & $34.3$ & $1180.9$ & $0.06$ & $5.52$ & $4.89$ & $1.01$ & TH & RCB\\
V* RY Sgr & $72.6$ & $252.5$ & $-55.1$ & $1772.2$ & $0.20$ & $9.64$ & $6.41$ & $1.21$ & TK & RCB\\
V* ES Aql & $-53.1$ & $180.3$ & $72.4$ & $536.9$ & $0.53$ & $5.33$ & $1.66$ & $1.22$ & TK & RCB\\
V* V482 Cyg & $-14.5$ & $215.0$ & $-18.3$ & $1595.6$ & $0.15$ & $7.95$ & $5.84$ & $0.36$ & TH & RCB\\
V* UV Cas & $52.4$ & $220.3$ & $4.5$ & $2143.8$ & $0.17$ & $11.17$ & $7.92$ & $0.06$ & TH & RCB\\
IRAS 00450+7401 & $40.0$ & $170.7$ & $-18.4$ & $3629.5$ & $0.20$ & $21.60$ & $14.50$ & $3.90$ & TK & RCB\\
V* W Men & $-53.1$ & $55.1$ & $-142.1$ & $425.4$ & $0.62$ & $9.60$ & $2.25$ & $7.98$ & H & RCB\\
HD 137613 & $-80.8$ & $197.6$ & $5.2$ & $1351.8$ & $0.30$ & $8.09$ & $4.37$ & $0.63$ & TH & HdC\\
C* 2277 & $-185.7$ & $91.5$ & $50.0$ & $536.6$ & $0.38$ & $4.14$ & $1.87$ & $0.90$ & H & HdC\\
SOPS   IV e-67 & $-17.7$ & $120.0$ & $-123.5$ & $638.4$ & $0.44$ & $6.25$ & $2.43$ & $3.20$ & H & HdC\\
V* LV TrA & $-33.4$ & $232.1$ & $40.3$ & $1504.5$ & $0.09$ & $7.01$ & $5.89$ & $0.61$ & TH & HdC\\
V* AC Ser & $-46.8$ & $156.2$ & $19.1$ & $-45.1$ & $0.88$ & $3.07$ & $0.20$ & $1.99$ & H & RCB\\
UCAC4 325-115052 & $-126.3$ & $283.5$ & $25.9$ & $2899.1$ & $0.50$ & $25.39$ & $8.46$ & $1.10$ & TK & RCB\\
2MASS J17485129-3306172 & $108.8$ & $336.9$ & $38.3$ & $3662.9$ & $0.60$ & $41.20$ & $10.14$ & $2.63$ & TK & RCB\\
2MASS J17510712-2423573 & $123.2$ & $-89.0$ & $54.8$ & $-98.5$ & $0.28$ & $1.05$ & $0.59$ & $0.23$ & B & RCB\\
IRAS 18029-1513 & $-192.8$ & $283.6$ & $33.0$ & $2015.0$ & $0.56$ & $18.85$ & $5.37$ & $0.79$ & H & RCB\\
ATO J277.4326-19.0462 & $-45.1$ & $266.9$ & $17.1$ & $2507.8$ & $0.21$ & $14.00$ & $9.15$ & $0.39$ & TH & RCB\\
HD 173409 & $109.3$ & $272.8$ & $-55.3$ & $4788.9$ & $0.45$ & $41.82$ & $16.05$ & $8.14$ & TK & HdC\\
V* V4152 Sgr & $-52.4$ & $172.9$ & $-34.0$ & $1024.8$ & $0.32$ & $6.48$ & $3.31$ & $0.81$ & TH & RCB\\
HD 182040 & $73.0$ & $236.6$ & $-7.8$ & $1810.5$ & $0.18$ & $9.33$ & $6.53$ & $0.21$ & TH & HdC\\
CD-35 13668 & $54.1$ & $360.8$ & $9.2$ & $4427.0$ & $0.66$ & $58.95$ & $12.18$ & $7.87$ & TK & HdC\\
C* 2891 & $37.0$ & $247.9$ & $-34.2$ & $2386.5$ & $0.07$ & $13.13$ & $11.35$ & $5.35$ & TK & HdC\\[2mm]
Ton S 137 & $-17.1$ & $141.7$ & $38.9$ & $1100.9$ & $0.44$ & $8.28$ & $3.23$ & $1.06$ & TK & He-sdO\\
PG 0016+151 & $-31.8$ & $239.8$ & $-28.7$ & $2113.6$ & $0.12$ & $10.91$ & $8.61$ & $2.21$ & TK & He-sdO\\
LAMOST J001954.03+510505.6 & $60.2$ & $226.3$ & $35.0$ & $1899.2$ & $0.26$ & $10.95$ & $6.50$ & $0.80$ & TH & He-sdO\\
PB  6221 & $-87.2$ & $122.3$ & $-8.9$ & $1271.5$ & $0.51$ & $13.30$ & $4.36$ & $6.67$ & H & He-sdO\\
LAMOST J010223.99+485245.3 & $-28.6$ & $220.1$ & $13.8$ & $2768.5$ & $0.17$ & $15.14$ & $10.78$ & $1.62$ & TK & He-sdO\\
HS 0110+3222 & $-8.4$ & $118.7$ & $-57.3$ & $1185.0$ & $0.50$ & $11.06$ & $3.68$ & $3.89$ & TK & He-sdO\\
PG 0113+259 & $59.3$ & $231.1$ & $-14.7$ & $2015.9$ & $0.20$ & $10.89$ & $7.33$ & $0.90$ & TH & He-sdO\\
PB  6433 & $46.0$ & $125.7$ & $-15.2$ & $1157.7$ & $0.49$ & $9.97$ & $3.45$ & $2.59$ & TK & He-sdO\\
GALEX J012630.2+503946 & $14.2$ & $253.2$ & $25.9$ & $2804.5$ & $0.17$ & $15.26$ & $10.93$ & $1.42$ & TK & He-sdO\\
FBS 0132+370 & $18.5$ & $265.3$ & $-36.3$ & $2442.2$ & $0.18$ & $13.32$ & $9.28$ & $1.44$ & TK & He-sdO\\
LAMOST J015012.51+254747.2 & $-99.9$ & $66.7$ & $76.5$ & $631.0$ & $0.77$ & $12.05$ & $1.59$ & $4.23$ & H & He-sdO\\
Feige  19 & $42.5$ & $253.3$ & $10.1$ & $2201.5$ & $0.16$ & $11.47$ & $8.38$ & $0.98$ & TH & He-sdO\\
\hline
\end{tabular}
\end{table*}
\begin{table*}

\setlength{\tabcolsep}{5pt}
\tabulinesep=1.5mm
\begin{tabular}{ L{4cm} rrrrr rrrcc}
\hline
Star & $U$ & $V$ & $W$ & $J_z$ & $e$ & R$_{a}$ & R$_p$ & Z$_{\rm max}$ &pop  &class \\
  & \kmsec & \kmsec & \kmsec & kpc \kmsec & & kpc & kpc & kpc &  &\\
\hline
FBS 0212+385 & $31.4$ & $262.3$ & $-39.6$ & $2192.0$ & $0.16$ & $11.50$ & $8.27$ & $0.91$ & TH & He-sdO\\
LAMOST J021703.37+544752.1 & $-51.1$ & $195.3$ & $17.6$ & $2086.1$ & $0.25$ & $11.93$ & $7.22$ & $0.52$ & TH & He-sdO\\
LAMOST J023851.64+432309.5 & $-22.7$ & $121.4$ & $1.4$ & $1146.3$ & $0.54$ & $10.11$ & $3.02$ & $0.55$ & TK & He-sdO\\
Feige  26 & $-26.6$ & $216.3$ & $99.0$ & $2036.4$ & $0.17$ & $11.86$ & $8.46$ & $4.10$ & TK & He-sdO\\
PG 0310+149 & $37.6$ & $246.0$ & $8.4$ & $2348.2$ & $0.13$ & $12.01$ & $9.27$ & $1.20$ & TH & He-sdO\\
UCAC4 551-007425 & $27.5$ & $203.3$ & $21.6$ & $2040.6$ & $0.11$ & $10.02$ & $8.11$ & $1.07$ & TH & He-sdO\\
EC 03505-6929 & $-41.8$ & $244.4$ & $51.0$ & $1921.5$ & $0.14$ & $9.99$ & $7.47$ & $1.65$ & TK & He-sdO\\
HE 0414-5429 & $-16.7$ & $255.6$ & $16.6$ & $2148.8$ & $0.13$ & $11.09$ & $8.50$ & $1.69$ & TK & He-sdO\\
GALEX J041536.0+253857 & $-65.2$ & $250.4$ & $31.8$ & $2524.7$ & $0.24$ & $14.87$ & $9.06$ & $1.29$ & TK & He-sdO\\
EC 04253-8213 & $-52.4$ & $252.6$ & $37.4$ & $1878.7$ & $0.18$ & $10.19$ & $7.07$ & $1.76$ & TK & He-sdO\\
GALEX J042542.2-060934 & $-12.0$ & $232.3$ & $-8.0$ & $2343.0$ & $0.05$ & $11.13$ & $10.08$ & $1.58$ & TK & He-sdO\\
BPS CS 22182-0037 & $1.8$ & $288.5$ & $39.1$ & $2554.4$ & $0.28$ & $15.91$ & $9.00$ & $2.05$ & TK & He-sdO\\
UCAC4 434-007650 & $-55.1$ & $277.4$ & $47.8$ & $2657.3$ & $0.32$ & $17.51$ & $9.11$ & $2.29$ & TK & He-sdO\\
GALEX J050231.9+091835 & $-11.0$ & $199.7$ & $-0.1$ & $2182.0$ & $0.15$ & $11.26$ & $8.34$ & $0.91$ & TH & He-sdO\\
GALEX J050325.3+641909 & $-32.1$ & $214.2$ & $-7.3$ & $2362.5$ & $0.14$ & $12.14$ & $9.20$ & $0.92$ & TH & He-sdO\\
GALEX J050650.6+193055 & $-44.5$ & $198.0$ & $20.2$ & $1980.6$ & $0.24$ & $11.27$ & $6.84$ & $0.62$ & TH & He-sdO\\
GALEX J051157.7+235610 & $-6.3$ & $172.0$ & $38.5$ & $2004.2$ & $0.24$ & $11.55$ & $7.12$ & $1.52$ & TK & He-sdO\\
UCAC4 535-014568 & $41.8$ & $218.6$ & $27.0$ & $2194.4$ & $0.05$ & $10.18$ & $9.17$ & $0.74$ & TH & He-sdO\\
UCAC4 630-027754 & $18.8$ & $191.9$ & $13.8$ & $1881.4$ & $0.16$ & $9.52$ & $6.96$ & $0.32$ & TH & He-sdO\\
KPD 0552+1903 & $-31.6$ & $180.9$ & $-15.1$ & $1849.1$ & $0.27$ & $10.74$ & $6.17$ & $0.29$ & TH & He-sdO\\
LAMOST J055814.17+464025.3 & $13.9$ & $231.4$ & $2.0$ & $2403.6$ & $0.03$ & $11.01$ & $10.28$ & $0.50$ & TH & He-sdO\\
UCAC4 498-020345 & $-110.5$ & $264.8$ & $27.7$ & $2580.6$ & $0.44$ & $20.08$ & $7.80$ & $1.03$ & TK & He-sdO\\
LAMOST J060806.00+461001.4 & $19.8$ & $190.8$ & $2.5$ & $2763.9$ & $0.12$ & $14.32$ & $11.18$ & $1.37$ & TK & He-sdO\\
GALEX J061205.9+473125 & $-1.0$ & $165.8$ & $30.7$ & $2105.3$ & $0.22$ & $12.07$ & $7.66$ & $1.83$ & TK & He-sdO\\
LAMOST J061924.34+234639.8 & $-43.4$ & $248.6$ & $-22.6$ & $2560.4$ & $0.08$ & $12.52$ & $10.56$ & $0.85$ & TH & He-sdO\\
UCAC4 555-028073 & $-1.6$ & $243.0$ & $18.7$ & $2803.0$ & $0.10$ & $14.08$ & $11.48$ & $0.69$ & TH & He-sdO\\
LAMOST J062812.03+374234.5 & $-21.0$ & $204.2$ & $-11.7$ & $2670.9$ & $0.13$ & $13.90$ & $10.64$ & $1.21$ & TK & He-sdO\\
KUV 06290+2813 & $25.1$ & $201.1$ & $41.7$ & $2145.4$ & $0.08$ & $10.36$ & $8.89$ & $1.41$ & TK & He-sdO\\
UCAC4 566-028577 & $-33.2$ & $139.4$ & $-50.0$ & $1339.2$ & $0.45$ & $10.31$ & $3.94$ & $1.65$ & TK & He-sdO\\
GALEX J063452.6+371110 & $37.5$ & $151.9$ & $6.2$ & $2584.1$ & $0.28$ & $16.06$ & $9.10$ & $1.95$ & TK & He-sdO\\
TYC 1337-283-1 & $-61.4$ & $253.2$ & $3.3$ & $2172.1$ & $0.23$ & $12.20$ & $7.62$ & $0.10$ & TH & He-sdO\\
FBS 0649+403 & $102.4$ & $210.8$ & $40.8$ & $2116.4$ & $0.28$ & $12.89$ & $7.25$ & $1.47$ & TK & He-sdO\\
SDSS J065256.41+120300.8 & $-76.0$ & $254.4$ & $-49.7$ & $3846.2$ & $0.12$ & $21.60$ & $17.03$ & $5.60$ & TK & He-sdO\\
HS 0657+5333 & $174.2$ & $22.7$ & $57.4$ & $583.3$ & $0.80$ & $12.53$ & $1.41$ & $3.18$ & H & He-sdO\\
2MASS J07024049+1147418 & $-8.0$ & $213.0$ & $4.1$ & $2309.7$ & $0.09$ & $11.18$ & $9.28$ & $0.44$ & TH & He-sdO\\
LAMOST J070801.51+103510.1 & $-4.9$ & $247.2$ & $3.9$ & $2795.5$ & $0.09$ & $13.86$ & $11.59$ & $0.72$ & TH & He-sdO\\
GALEX J071503.2+750140 & $-119.6$ & $-106.2$ & $-21.6$ & $-671.8$ & $0.85$ & $18.55$ & $1.55$ & $4.33$ & H & He-sdO\\
GALEX J072516.9+283831 & $8.6$ & $263.1$ & $18.9$ & $2391.6$ & $0.13$ & $12.20$ & $9.31$ & $0.64$ & TH & He-sdO\\
LAMOST J072628.27+083747.9 & $-105.9$ & $104.5$ & $-72.4$ & $1413.5$ & $0.60$ & $17.36$ & $4.32$ & $7.88$ & H & He-sdO\\
LAMOST J072835.11+280239.1 & $2.0$ & $232.7$ & $15.1$ & $2604.5$ & $0.04$ & $12.24$ & $11.39$ & $1.34$ & TK & He-sdO\\
UCAC4 453-037143 & $-15.8$ & $198.9$ & $-28.7$ & $2019.5$ & $0.33$ & $13.51$ & $6.79$ & $2.69$ & TK & He-sdO\\
LAMOST J073409.31+523028.6 & $-22.0$ & $-92.3$ & $-3.5$ & $-278.4$ & $0.91$ & $12.27$ & $0.58$ & $1.56$ & H & He-sdO\\
HS 0735+4026 & $22.5$ & $171.8$ & $22.9$ & $2192.6$ & $0.17$ & $11.93$ & $8.53$ & $2.24$ & TK & He-sdO\\
HS 0736+3953 & $38.6$ & $123.5$ & $32.0$ & $2758.5$ & $0.26$ & $18.80$ & $11.02$ & $7.14$ & TK & He-sdO\\
FBS 0742+337 & $-54.6$ & $275.9$ & $-30.3$ & $2641.5$ & $0.11$ & $13.74$ & $11.03$ & $2.48$ & TK & He-sdO\\
GALEX J075234.2+161604 & $-34.7$ & $244.2$ & $-2.7$ & $2220.4$ & $0.07$ & $10.50$ & $9.08$ & $0.60$ & TH & He-sdO\\
GALEX J075250.0+305936 & $40.4$ & $25.9$ & $11.5$ & $755.8$ & $0.75$ & $12.84$ & $1.85$ & $3.32$ & TK & He-sdO\\
BD+75   325 & $-22.6$ & $243.7$ & $0.8$ & $2018.3$ & $0.10$ & $9.60$ & $7.92$ & $0.11$ & TH & He-sdO\\
GALEX J082751.0+410925 & $-17.7$ & $322.1$ & $12.8$ & $1759.6$ & $0.39$ & $13.25$ & $5.85$ & $3.90$ & TK & He-sdO\\
$\rm [CW83]\, 0832-01$ & $10.8$ & $235.9$ & $26.3$ & $2109.4$ & $0.11$ & $10.35$ & $8.29$ & $0.68$ & TH & He-sdO\\
PG 0838+133 & $-69.2$ & $253.4$ & $-22.5$ & $2230.2$ & $0.19$ & $12.10$ & $8.28$ & $1.13$ & TH & He-sdO\\
US 1993 & $4.1$ & $58.9$ & $9.3$ & $609.0$ & $0.76$ & $10.65$ & $1.46$ & $1.97$ & TK & He-sdO\\
GALEX J090253.0+073533 & $25.4$ & $-32.3$ & $-17.4$ & $306.3$ & $0.91$ & $16.08$ & $0.75$ & $6.47$ & H & He-sdO\\
$\rm [CW83]\, 0904-02$ & $-84.0$ & $315.3$ & $-18.1$ & $2189.9$ & $0.17$ & $11.69$ & $8.26$ & $1.24$ & TK & He-sdO\\
PG 0912+119 & $-71.9$ & $118.9$ & $-16.5$ & $924.8$ & $0.62$ & $10.34$ & $2.45$ & $2.21$ & TK & He-sdO\\
PG 0914-037 & $-166.9$ & $152.7$ & $-71.1$ & $971.9$ & $0.47$ & $9.93$ & $3.55$ & $5.26$ & H & He-sdO\\
Ton  414 & $-4.0$ & $176.0$ & $4.5$ & $1850.6$ & $0.25$ & $10.77$ & $6.40$ & $1.26$ & TK & He-sdO\\
SBSS 0934+495 & $13.5$ & $137.4$ & $31.6$ & $1715.3$ & $0.33$ & $12.29$ & $6.17$ & $4.39$ & TK & He-sdO\\
TYC 4895-599-1 & $-98.9$ & $292.6$ & $28.9$ & $1856.0$ & $0.35$ & $12.25$ & $5.93$ & $1.37$ & TK & He-sdO\\
GD 300 & $-109.9$ & $99.8$ & $-0.1$ & $-405.3$ & $0.84$ & $11.18$ & $0.96$ & $2.55$ & TK & He-sdO\\
GALEX J095601.7+091138 & $-54.8$ & $217.1$ & $-7.9$ & $1887.0$ & $0.17$ & $10.00$ & $7.06$ & $1.33$ & TK & He-sdO\\
GALEX J095648.5+142241 & $-124.9$ & $265.4$ & $-48.6$ & $1922.5$ & $0.06$ & $9.53$ & $8.52$ & $2.68$ & TK & He-sdO\\
Ton  1137 & $-73.5$ & $204.0$ & $-4.3$ & $1463.8$ & $0.36$ & $10.25$ & $4.84$ & $2.65$ & TK & He-sdO\\
PG 1034+001 & $-51.8$ & $277.8$ & $59.9$ & $1725.4$ & $0.28$ & $10.15$ & $5.70$ & $0.54$ & TH & He-sdO\\
GALEX J104033.8+562206 & $-7.5$ & $188.4$ & $33.6$ & $1665.4$ & $0.26$ & $10.20$ & $5.96$ & $2.49$ & TK & He-sdO\\
PG 1038+510 & $-128.7$ & $205.8$ & $108.0$ & $-250.5$ & $0.95$ & $21.02$ & $0.50$ & $3.76$ & TK & He-sdO\\

\hline
\end{tabular}
\end{table*}

\begin{table*}

\setlength{\tabcolsep}{5pt}
\tabulinesep=1.5mm
\begin{tabular}{ L{4cm} rrrrr rrrcc}
\hline
Star & $U$ & $V$ & $W$ & $J_z$ & $e$ & R$_{a}$ & R$_p$ & Z$_{\rm max}$ &pop  &class \\
  & \kmsec & \kmsec & \kmsec & kpc \kmsec & & kpc & kpc & kpc &  &\\
\hline
GALEX J105759.3+470307 & $-1.0$ & $215.2$ & $16.4$ & $2121.7$ & $0.05$ & $10.45$ & $9.46$ & $2.76$ & TK & He-sdO\\
SDSS J110215.45+024034.1 & $-138.1$ & $100.0$ & $-60.3$ & $832.6$ & $0.68$ & $12.44$ & $2.34$ & $5.17$ & H & He-sdO\\
PG 1102+499 & $-53.9$ & $210.7$ & $-19.6$ & $1627.2$ & $0.23$ & $9.37$ & $5.91$ & $1.96$ & TK & He-sdO\\
CBS 432 & $-95.0$ & $193.2$ & $29.6$ & $1141.1$ & $0.50$ & $10.90$ & $3.62$ & $4.20$ & TK & He-sdO\\
PG 1125-055 & $-224.4$ & $5.5$ & $172.6$ & $-782.5$ & $0.61$ & $10.10$ & $2.48$ & $4.83$ & H & He-sdO\\
PG 1134+463 & $-88.8$ & $227.3$ & $35.2$ & $1216.1$ & $0.46$ & $9.79$ & $3.59$ & $1.97$ & TK & He-sdO\\
HS 1203+6650 & $-17.9$ & $249.0$ & $41.6$ & $1496.1$ & $0.45$ & $11.99$ & $4.57$ & $2.95$ & TK & He-sdO\\
GALEX J123808.7+053318 & $-191.3$ & $173.9$ & $5.3$ & $830.5$ & $0.51$ & $8.66$ & $2.82$ & $4.08$ & H & He-sdO\\
$\rm [BFS85]\, 12h30\,  2B1$ & $-100.4$ & $209.5$ & $-4.7$ & $1499.3$ & $0.24$ & $10.23$ & $6.21$ & $4.75$ & H & He-sdO\\
Ton  102 & $-11.2$ & $208.5$ & $93.8$ & $1857.0$ & $0.11$ & $10.04$ & $8.03$ & $3.40$ & TK & He-sdO\\
Ton  109 & $-233.6$ & $126.0$ & $-94.8$ & $240.0$ & $0.56$ & $9.33$ & $2.63$ & $8.93$ & H & He-sdO\\
PG 1249+762 & $-9.1$ & $232.3$ & $50.7$ & $1541.8$ & $0.37$ & $10.58$ & $4.86$ & $1.75$ & TK & He-sdO\\
SDSS J125301.62+394622.2 & $-86.0$ & $64.9$ & $-111.5$ & $905.1$ & $0.71$ & $14.29$ & $2.39$ & $5.30$ & H & He-sdO\\
PG 1251+019 & $-109.8$ & $-11.9$ & $-143.2$ & $118.4$ & $0.97$ & $15.16$ & $0.27$ & $6.56$ & H & He-sdO\\
PB  4363 & $-306.5$ & $207.1$ & $81.0$ & $837.0$ & $0.46$ & $10.95$ & $4.04$ & $8.13$ & H & He-sdO\\
Ton  143 & $-108.5$ & $360.2$ & $83.2$ & $1658.6$ & $0.55$ & $17.39$ & $5.00$ & $6.37$ & H & He-sdO\\
PG 1316+212 & $-249.7$ & $106.6$ & $-18.8$ & $-165.9$ & $0.88$ & $8.40$ & $0.53$ & $4.58$ & H & He-sdO\\
PG 1318+062 & $-110.3$ & $304.4$ & $86.9$ & $1636.3$ & $0.31$ & $10.48$ & $5.57$ & $2.27$ & TK & He-sdO\\
PB   166 & $-36.8$ & $267.3$ & $-0.5$ & $1694.9$ & $0.25$ & $9.75$ & $5.81$ & $1.07$ & TH & He-sdO\\
PG 1325+054 & $-48.9$ & $310.7$ & $37.5$ & $1967.9$ & $0.25$ & $11.56$ & $6.97$ & $1.79$ & TK & He-sdO\\
GALEX J134738.3+043427 & $-165.8$ & $181.7$ & $29.6$ & $767.8$ & $0.40$ & $8.00$ & $3.44$ & $5.26$ & H & He-sdO\\
CD-46  8926 & $-3.0$ & $321.3$ & $-45.4$ & $2035.4$ & $0.43$ & $15.59$ & $6.28$ & $2.77$ & TK & He-sdO\\
SDSS J135707.35+010454.4 & $-279.3$ & $115.0$ & $49.2$ & $81.3$ & $0.90$ & $7.67$ & $0.41$ & $5.74$ & H & He-sdO\\
PG 1401+289 & $-94.2$ & $150.3$ & $-64.2$ & $875.9$ & $0.50$ & $8.52$ & $2.81$ & $3.32$ & TK & He-sdO\\
GALEX J140715.4+033147 & $-191.9$ & $207.7$ & $73.5$ & $678.0$ & $0.51$ & $7.56$ & $2.45$ & $3.95$ & H & He-sdO\\
PG 1412+613 & $-20.8$ & $202.4$ & $-20.0$ & $1692.1$ & $0.17$ & $8.97$ & $6.42$ & $1.61$ & TK & He-sdO\\
FBS 1412+004 & $-173.7$ & $111.0$ & $-12.9$ & $377.6$ & $0.69$ & $7.61$ & $1.39$ & $4.87$ & H & He-sdO\\
PG 1427+196 & $-43.2$ & $228.1$ & $17.2$ & $1609.9$ & $0.11$ & $7.97$ & $6.37$ & $1.36$ & TK & He-sdO\\
PG 1442+346 & $-23.0$ & $173.7$ & $-90.5$ & $945.6$ & $0.32$ & $8.26$ & $4.27$ & $5.12$ & H & He-sdO\\
PG 1444+076 & $-414.9$ & $31.7$ & $80.7$ & $-1228.6$ & $0.32$ & $7.83$ & $4.08$ & $1.50$ & H & He-sdO\\
GALEX J144738.3+615033 & $-10.1$ & $232.4$ & $77.2$ & $1724.5$ & $0.22$ & $11.22$ & $7.11$ & $4.82$ & TK & He-sdO\\
GALEX J145134.2+252221 & $18.3$ & $207.9$ & $-91.2$ & $1389.4$ & $0.19$ & $8.58$ & $5.87$ & $3.67$ & TK & He-sdO\\
PG 1502+129 & $-8.5$ & $270.8$ & $22.6$ & $1284.1$ & $0.38$ & $9.71$ & $4.37$ & $3.31$ & TK & He-sdO\\
PG 1507-015 & $-63.9$ & $259.7$ & $62.1$ & $1401.8$ & $0.12$ & $7.28$ & $5.72$ & $1.96$ & TK & He-sdO\\
GALEX J151415.6-012925 & $-164.2$ & $180.1$ & $56.2$ & $587.4$ & $0.60$ & $6.79$ & $1.72$ & $1.83$ & TK & He-sdO\\
PG 1528+029 & $-87.7$ & $229.7$ & $56.9$ & $1141.2$ & $0.26$ & $6.96$ & $4.08$ & $1.75$ & TK & He-sdO\\
PG 1534-018 & $66.6$ & $252.0$ & $-53.4$ & $1957.4$ & $0.26$ & $11.91$ & $6.92$ & $2.28$ & TK & He-sdO\\
PG 1539+442 & $-118.9$ & $218.6$ & $72.4$ & $1315.3$ & $0.24$ & $9.33$ & $5.69$ & $4.84$ & H & He-sdO\\
PG 1539+043 & $-180.0$ & $254.6$ & $137.5$ & $-92.0$ & $0.81$ & $6.65$ & $0.69$ & $6.01$ & H & He-sdO\\
SDSS J154801.15-023452.3 & $-130.2$ & $130.0$ & $16.7$ & $-256.4$ & $0.47$ & $6.30$ & $2.27$ & $5.79$ & H & He-sdO\\
PG 1554+222 & $-11.1$ & $221.5$ & $13.9$ & $1288.8$ & $0.32$ & $8.10$ & $4.21$ & $1.25$ & TK & He-sdO\\
PG 1555+504 & $28.5$ & $254.5$ & $99.6$ & $1344.0$ & $0.49$ & $11.86$ & $4.11$ & $3.50$ & TK & He-sdO\\
PG 1555+489 & $-54.1$ & $222.6$ & $76.6$ & $1202.2$ & $0.34$ & $8.63$ & $4.25$ & $3.03$ & TK & He-sdO\\
GALEX J160053.0-044132 & $-98.6$ & $195.8$ & $44.1$ & $874.7$ & $0.47$ & $7.11$ & $2.53$ & $0.95$ & TK & He-sdO\\
GALEX J160152.8-040949 & $84.3$ & $133.8$ & $-205.8$ & $1089.9$ & $0.53$ & $12.63$ & $3.85$ & $7.02$ & H & He-sdO\\
PG 1607+228 & $-37.3$ & $224.1$ & $25.7$ & $1299.7$ & $0.13$ & $7.08$ & $5.41$ & $2.26$ & TK & He-sdO\\
PG 1610+043 & $-65.6$ & $227.4$ & $44.9$ & $1157.1$ & $0.19$ & $6.53$ & $4.40$ & $1.70$ & TK & He-sdO\\
PG 1611+041 & $-136.3$ & $176.0$ & $24.4$ & $792.2$ & $0.47$ & $6.88$ & $2.46$ & $1.84$ & TK & He-sdO\\
GALEX J161627.1-002933 & $-81.1$ & $217.9$ & $32.2$ & $1101.4$ & $0.20$ & $6.38$ & $4.22$ & $1.84$ & TK & He-sdO\\
Ton  257 & $-182.5$ & $172.3$ & $97.8$ & $113.4$ & $0.92$ & $7.20$ & $0.31$ & $3.35$ & H & He-sdO\\
GALEX J162411.5+312252 & $74.4$ & $166.3$ & $80.7$ & $-234.8$ & $0.87$ & $11.16$ & $0.81$ & $8.21$ & H & He-sdO\\
PG 1624+382 & $-27.9$ & $241.8$ & $53.5$ & $1561.5$ & $0.12$ & $7.99$ & $6.24$ & $1.78$ & TK & He-sdO\\
GALEX J162626.2+333314 & $19.7$ & $231.4$ & $81.6$ & $920.2$ & $0.51$ & $9.13$ & $2.96$ & $3.70$ & TK & He-sdO\\
PG 1624+085 & $-37.5$ & $198.8$ & $-2.4$ & $1224.5$ & $0.27$ & $7.22$ & $4.11$ & $0.96$ & TH & He-sdO\\
PG 1625-034 & $42.3$ & $232.6$ & $6.3$ & $1342.2$ & $0.32$ & $8.37$ & $4.35$ & $1.09$ & TH & He-sdO\\
GALEX J170045.5+604307 & $49.1$ & $246.5$ & $82.8$ & $2065.0$ & $0.30$ & $15.11$ & $8.10$ & $6.62$ & TK & He-sdO\\
PG 1700+198 & $-88.5$ & $240.3$ & $103.5$ & $1124.4$ & $0.26$ & $7.13$ & $4.15$ & $2.27$ & TK & He-sdO\\
PG 1703+355 & $2.5$ & $241.8$ & $89.4$ & $1356.4$ & $0.32$ & $8.92$ & $4.63$ & $2.26$ & TK & He-sdO\\
GALEX J171720.5+094131 & $-37.6$ & $126.5$ & $-43.5$ & $576.8$ & $0.57$ & $6.33$ & $1.73$ & $1.70$ & TK & He-sdO\\
GALEX J174516.3+244348 & $-76.3$ & $120.9$ & $-109.2$ & $717.6$ & $0.53$ & $8.10$ & $2.49$ & $4.01$ & H & He-sdO\\
GALEX J175548.5+501210 & $63.6$ & $200.6$ & $8.0$ & $1549.3$ & $0.30$ & $9.29$ & $4.96$ & $0.33$ & TH & He-sdO\\
LAMOST J181712.86+053202.7 & $-41.7$ & $205.8$ & $12.5$ & $1010.3$ & $0.27$ & $5.92$ & $3.39$ & $0.61$ & TH & He-sdO\\
GALEX J182208.6+103745 & $72.0$ & $281.3$ & $17.8$ & $1217.4$ & $0.18$ & $6.79$ & $4.71$ & $1.85$ & TK & He-sdO\\
UCAC4 687-068720 & $14.2$ & $277.3$ & $9.7$ & $2210.4$ & $0.19$ & $11.90$ & $8.16$ & $0.82$ & TH & He-sdO\\
LAMOST J193312.83+380154.6 & $-48.6$ & $237.2$ & $21.4$ & $1670.1$ & $0.16$ & $8.54$ & $6.17$ & $0.80$ & TH & He-sdO\\
BPS CS 22896-0128 & $43.1$ & $213.0$ & $42.1$ & $1115.5$ & $0.12$ & $6.01$ & $4.69$ & $1.98$ & TK & He-sdO\\
\hline
\end{tabular}
\end{table*}

\begin{table*}

\setlength{\tabcolsep}{5pt}
\tabulinesep=1.5mm
\begin{tabular}{ L{4cm} rrrrr rrrcc}
\hline
Star & $U$ & $V$ & $W$ & $J_z$ & $e$ & R$_{a}$ & R$_p$ & Z$_{\rm max}$ &pop  &class \\
  & \kmsec & \kmsec & \kmsec & kpc \kmsec & & kpc & kpc & kpc &  &\\
\hline

LAMOST J202546.17+092158.7 & $-29.2$ & $221.5$ & $56.6$ & $1173.3$ & $0.23$ & $6.97$ & $4.34$ & $1.88$ & TK & He-sdO\\
LAMOST J202610.56+055413.9 & $22.0$ & $194.3$ & $-8.6$ & $1361.8$ & $0.21$ & $7.28$ & $4.77$ & $0.65$ & TH & He-sdO\\
LAMOST J210936.02+362337.5 & $-41.3$ & $227.9$ & $4.5$ & $1892.2$ & $0.15$ & $9.56$ & $7.08$ & $0.56$ & TH & He-sdO\\
HS 2108+1413 & $-39.0$ & $273.8$ & $-15.9$ & $2108.2$ & $0.20$ & $11.41$ & $7.64$ & $0.66$ & TH & He-sdO\\
PG 2110+001 & $-103.0$ & $206.2$ & $25.2$ & $1298.7$ & $0.34$ & $8.62$ & $4.24$ & $1.73$ & TK & He-sdO\\
LAMOST J211624.57+132338.7 & $99.8$ & $19.5$ & $28.2$ & $209.4$ & $0.88$ & $11.70$ & $0.73$ & $8.23$ & H & He-sdO\\
GALEX J211832.4+233416 & $-48.1$ & $218.1$ & $9.3$ & $1652.4$ & $0.18$ & $8.60$ & $5.95$ & $0.63$ & TH & He-sdO\\
PG 2116+008 & $35.9$ & $224.3$ & $-8.2$ & $1568.9$ & $0.11$ & $7.86$ & $6.25$ & $1.54$ & TK & He-sdO\\
GALEX J211925.9+202155 & $24.5$ & $128.7$ & $-73.7$ & $1014.1$ & $0.42$ & $8.00$ & $3.28$ & $2.40$ & TK & He-sdO\\
GALEX J212356.6+153323 & $63.0$ & $90.5$ & $-42.9$ & $718.7$ & $0.61$ & $8.05$ & $1.94$ & $1.56$ & TK & He-sdO\\
GALEX J212504.4+212957 & $13.8$ & $244.6$ & $21.9$ & $1868.4$ & $0.05$ & $8.78$ & $8.00$ & $1.66$ & TK & He-sdO\\
GALEX J212544.4+202812 & $-61.1$ & $237.9$ & $9.7$ & $1844.4$ & $0.18$ & $9.64$ & $6.64$ & $0.38$ & TH & He-sdO\\
UCAC4 648-107111 & $-22.2$ & $254.9$ & $-5.9$ & $2089.3$ & $0.12$ & $10.26$ & $8.07$ & $0.24$ & TH & He-sdO\\
GALEX J214106.6+252832 & $-92.7$ & $239.8$ & $0.3$ & $1886.5$ & $0.29$ & $11.25$ & $6.23$ & $0.47$ & TH & He-sdO\\
SDSS J214226.13+091307.5 & $140.4$ & $50.0$ & $53.4$ & $408.0$ & $0.78$ & $9.03$ & $1.09$ & $3.73$ & H & He-sdO\\
GALEX J214540.8-012807 & $2.1$ & $83.9$ & $7.4$ & $589.6$ & $0.65$ & $7.54$ & $1.59$ & $1.46$ & TK & He-sdO\\
UCAC4 637-113968 & $4.7$ & $314.8$ & $-50.9$ & $2573.2$ & $0.37$ & $18.22$ & $8.31$ & $1.89$ & TK & He-sdO\\
PHL   149 & $-72.0$ & $179.7$ & $32.1$ & $1282.8$ & $0.35$ & $8.39$ & $4.08$ & $1.23$ & TK & He-sdO\\
PG 2158+082 & $-196.5$ & $260.6$ & $-72.6$ & $1998.7$ & $0.58$ & $20.97$ & $5.55$ & $4.84$ & H & He-sdO\\
PG 2201+145 & $-7.2$ & $227.7$ & $-34.7$ & $1786.7$ & $0.04$ & $8.14$ & $7.58$ & $1.18$ & TH & He-sdO\\
SDSS J220819.50+060255.4 & $-73.3$ & $183.2$ & $26.8$ & $1310.5$ & $0.33$ & $9.48$ & $4.73$ & $3.58$ & TK & He-sdO\\
BPS CS 22956-0090 & $16.3$ & $120.0$ & $69.5$ & $817.5$ & $0.52$ & $7.68$ & $2.45$ & $2.09$ & TK & He-sdO\\
BPS CS 22892-0051 & $151.3$ & $134.4$ & $94.6$ & $1000.4$ & $0.51$ & $10.38$ & $3.39$ & $4.99$ & H & He-sdO\\
BPS CS 22881-0069 & $48.7$ & $256.4$ & $26.5$ & $1927.0$ & $0.18$ & $10.33$ & $7.11$ & $1.21$ & TK & He-sdO\\
GALEX J223455.5+312325 & $56.9$ & $182.7$ & $63.9$ & $1592.0$ & $0.23$ & $9.49$ & $5.93$ & $2.60$ & TK & He-sdO\\
PHL   364 & $-7.8$ & $167.8$ & $-49.2$ & $1291.0$ & $0.28$ & $8.05$ & $4.54$ & $2.03$ & TK & He-sdO\\
PG 2244+153 & $130.3$ & $120.4$ & $26.9$ & $963.2$ & $0.60$ & $10.14$ & $2.54$ & $1.82$ & TK & He-sdO\\
PG 2249+220 & $11.7$ & $188.0$ & $11.7$ & $1605.4$ & $0.20$ & $8.90$ & $5.93$ & $1.76$ & TK & He-sdO\\
BPS CS 22938-0044 & $20.3$ & $230.2$ & $52.2$ & $1531.2$ & $0.15$ & $8.26$ & $6.12$ & $2.28$ & TK & He-sdO\\
GALEX J230124.7+325943 & $50.9$ & $87.8$ & $37.9$ & $831.7$ & $0.66$ & $11.02$ & $2.27$ & $3.69$ & TK & He-sdO\\
FBS 2307+338 & $10.8$ & $213.8$ & $30.8$ & $1864.7$ & $0.08$ & $8.88$ & $7.56$ & $1.11$ & TH & He-sdO\\
BPS CS 22938-0073 & $-33.5$ & $128.8$ & $17.8$ & $865.0$ & $0.53$ & $7.73$ & $2.41$ & $1.07$ & TK & He-sdO\\
PG 2333-002 & $76.1$ & $137.3$ & $74.5$ & $1114.1$ & $0.35$ & $9.40$ & $4.53$ & $5.04$ & H & He-sdO\\
GALEX J233913.9+134215 & $117.1$ & $113.0$ & $-84.8$ & $960.5$ & $0.63$ & $12.31$ & $2.76$ & $4.83$ & TK & He-sdO\\
PG 2343+267 & $26.2$ & $224.4$ & $13.7$ & $2145.5$ & $0.06$ & $10.48$ & $9.31$ & $2.30$ & TK & He-sdO\\
GALEX J235050.1+194628 & $56.8$ & $205.0$ & $47.6$ & $2056.4$ & $0.13$ & $11.87$ & $9.22$ & $4.85$ & TK & He-sdO\\
PG 2352+181 & $105.6$ & $190.9$ & $-22.2$ & $1548.4$ & $0.40$ & $10.84$ & $4.63$ & $0.80$ & TK & He-sdO\\[2mm]
GALEX J014640.5+411826 & $23.8$ & $264.3$ & $9.3$ & $2746.5$ & $0.19$ & $15.39$ & $10.44$ & $1.51$ & TK & He-sdB\\
UCAC4 450-010156 & $8.1$ & $243.1$ & $5.8$ & $2219.6$ & $0.05$ & $10.18$ & $9.26$ & $0.38$ & TH & He-sdB\\
GALEX J062038.5-570538 & $-16.2$ & $272.9$ & $5.9$ & $2229.0$ & $0.16$ & $11.56$ & $8.44$ & $0.64$ & TH & He-sdB\\
GALEX J064024.8+444720 & $-3.1$ & $249.6$ & $13.4$ & $2474.0$ & $0.10$ & $12.26$ & $9.99$ & $0.75$ & TH & He-sdB\\
GALEX J064918.3+235440 & $34.5$ & $226.9$ & $25.5$ & $2495.9$ & $0.07$ & $11.96$ & $10.47$ & $0.97$ & TH & He-sdB\\
FBS 0654+366 & $-3.5$ & $286.1$ & $-4.2$ & $2391.1$ & $0.19$ & $12.92$ & $8.85$ & $0.45$ & TH & He-sdB\\
HZ 38 & $-83.2$ & $264.7$ & $-21.6$ & $1538.0$ & $0.25$ & $9.05$ & $5.44$ & $1.80$ & TK & He-sdB\\
GALEX J145817.5+022807 & $-36.9$ & $240.8$ & $16.9$ & $874.4$ & $0.11$ & $7.86$ & $6.32$ & $6.29$ & H & He-sdB\\
PG 1607+174 & $-39.0$ & $228.0$ & $12.4$ & $1684.5$ & $0.12$ & $8.13$ & $6.41$ & $0.48$ & TH & He-sdB\\
FBS 1749+373 & $21.4$ & $193.2$ & $4.7$ & $1416.2$ & $0.28$ & $8.25$ & $4.61$ & $0.40$ & TH & He-sdB\\
UCAC4 657-066324 & $3.4$ & $232.4$ & $3.8$ & $1791.1$ & $0.04$ & $7.99$ & $7.37$ & $0.46$ & TH & He-sdB\\
2MASS J19230068+3715044 & $-68.6$ & $247.0$ & $-32.9$ & $1832.1$ & $0.22$ & $10.12$ & $6.49$ & $0.98$ & TH & He-sdB\\
GALEX J213308.4+064701 & $-76.8$ & $287.1$ & $52.6$ & $2991.3$ & $0.39$ & $23.22$ & $10.31$ & $6.43$ & TK & He-sdB\\
UCAC4 653-107038 & $-29.6$ & $212.6$ & $-5.9$ & $1817.2$ & $0.15$ & $9.14$ & $6.74$ & $0.46$ & TH & He-sdB\\
2MASS J21391157+4539155 & $-72.3$ & $182.6$ & $-17.8$ & $1815.0$ & $0.32$ & $11.38$ & $5.82$ & $0.70$ & TH & He-sdB\\
JL  87 & $-24.2$ & $264.9$ & $8.7$ & $2043.0$ & $0.15$ & $10.45$ & $7.67$ & $0.59$ & TH & He-sdB\\
BPS CS 22956-0094 & $63.8$ & $184.2$ & $21.7$ & $1232.8$ & $0.48$ & $9.66$ & $3.41$ & $0.51$ & TH & He-sdB\\[2mm]
UCAC4 685-001272 & $-4.8$ & $231.1$ & $26.0$ & $2073.9$ & $0.05$ & $9.48$ & $8.64$ & $0.64$ & TH & He-sdOB\\
GALEX J001306.4+491149 & $114.6$ & $58.2$ & $29.8$ & $-20.8$ & $0.98$ & $12.99$ & $0.11$ & $1.44$ & H & He-sdOB\\
GALEX J002433.0+264910 & $-4.2$ & $226.2$ & $16.2$ & $2145.4$ & $0.04$ & $10.12$ & $9.34$ & $1.82$ & TK & He-sdOB\\
FBS 0035+343 & $35.7$ & $302.5$ & $-41.4$ & $2604.3$ & $0.35$ & $17.68$ & $8.61$ & $1.67$ & TK & He-sdOB\\
PG 0039+135 & $-22.8$ & $191.0$ & $72.8$ & $1607.6$ & $0.19$ & $8.96$ & $6.08$ & $2.10$ & TK & He-sdOB\\
PB  6244 & $188.0$ & $102.9$ & $55.4$ & $994.0$ & $0.63$ & $15.49$ & $3.53$ & $9.76$ & H & He-sdOB\\
PB  6275 & $163.0$ & $-22.4$ & $-116.1$ & $-219.4$ & $0.92$ & $13.70$ & $0.58$ & $7.41$ & H & He-sdOB\\
GALEX J011928.8+490109 & $13.1$ & $239.5$ & $-9.6$ & $2048.5$ & $0.05$ & $9.28$ & $8.46$ & $0.25$ & TH & He-sdOB\\
GALEX J014326.2+323439 & $-11.7$ & $50.0$ & $68.6$ & $385.4$ & $0.82$ & $9.94$ & $0.95$ & $3.72$ & TK & He-sdOB\\
PG 0216+246 & $-38.3$ & $250.8$ & $8.7$ & $2480.3$ & $0.16$ & $13.40$ & $9.64$ & $1.79$ & TK & He-sdOB\\
GALEX J022125.9+085919 & $-86.3$ & $161.6$ & $-14.2$ & $1884.2$ & $0.37$ & $14.76$ & $6.76$ & $5.92$ & TK & He-sdOB\\
\hline
\end{tabular}
\end{table*}

\begin{table*}

\setlength{\tabcolsep}{5pt}
\tabulinesep=1.5mm
\begin{tabular}{ L{4cm} rrrrr rrrcc}
\hline
Star & $U$ & $V$ & $W$ & $J_z$ & $e$ & R$_{a}$ & R$_p$ & Z$_{\rm max}$ &pop  &class \\
  & \kmsec & \kmsec & \kmsec & kpc \kmsec & & kpc & kpc & kpc &  &\\
\hline

PG 0221+217 & $-122.6$ & $151.6$ & $-16.9$ & $1336.9$ & $0.52$ & $11.52$ & $3.65$ & $1.26$ & TK & He-sdOB\\
GALEX J022422.2+000313 & $21.2$ & $-33.8$ & $-39.2$ & $-297.3$ & $0.89$ & $12.93$ & $0.78$ & $6.54$ & H & He-sdOB\\
GALEX J023254.0+370419 & $-75.1$ & $218.0$ & $1.1$ & $1949.4$ & $0.26$ & $11.23$ & $6.62$ & $0.45$ & TH & He-sdOB\\
PG 0240+046 & $32.6$ & $164.2$ & $93.0$ & $1482.0$ & $0.20$ & $8.84$ & $5.87$ & $3.04$ & TK & He-sdOB\\
KUV 02445+3633 & $21.1$ & $295.9$ & $-90.1$ & $2562.5$ & $0.35$ & $18.10$ & $8.79$ & $4.03$ & TK & He-sdOB\\
GALEX J031842.4+340554 & $-19.9$ & $202.6$ & $9.7$ & $1875.7$ & $0.17$ & $9.65$ & $6.87$ & $0.45$ & TH & He-sdOB\\
KPD 0319+4553 & $78.9$ & $255.8$ & $-12.8$ & $2293.9$ & $0.27$ & $13.59$ & $7.87$ & $0.33$ & TH & He-sdOB\\
HD 275347 & $-35.6$ & $230.9$ & $2.8$ & $1998.0$ & $0.12$ & $9.77$ & $7.65$ & $0.13$ & TH & He-sdOB\\
BPS CS 22190-0003 & $-73.8$ & $277.7$ & $-6.5$ & $2399.4$ & $0.26$ & $14.49$ & $8.42$ & $1.49$ & TK & He-sdOB\\
HZ  3 & $-18.8$ & $232.2$ & $35.4$ & $2006.8$ & $0.10$ & $9.65$ & $7.97$ & $0.74$ & TH & He-sdOB\\
GALEX J041700.7+350717 & $34.4$ & $239.0$ & $12.2$ & $2356.8$ & $0.10$ & $11.51$ & $9.44$ & $0.45$ & TH & He-sdOB\\
UCAC4 633-017614 & $31.3$ & $167.9$ & $17.5$ & $1578.5$ & $0.28$ & $9.14$ & $5.18$ & $0.39$ & TH & He-sdOB\\
KUV 04456+1502 & $22.4$ & $203.2$ & $16.9$ & $1990.2$ & $0.10$ & $9.58$ & $7.83$ & $0.60$ & TH & He-sdOB\\
GALEX J045027.6+155910 & $32.5$ & $218.7$ & $5.1$ & $2216.0$ & $0.07$ & $10.45$ & $9.08$ & $0.59$ & TH & He-sdOB\\
GALEX J052300.5+182727 & $43.7$ & $196.7$ & $14.0$ & $2004.2$ & $0.13$ & $10.00$ & $7.62$ & $0.46$ & TH & He-sdOB\\
SDSS J052933.85+040720.0 & $-98.9$ & $186.6$ & $-11.1$ & $1837.3$ & $0.39$ & $12.77$ & $5.57$ & $0.74$ & TK & He-sdOB\\
GALEX J053656.5+395516 & $15.6$ & $203.5$ & $12.9$ & $1868.9$ & $0.12$ & $9.05$ & $7.14$ & $0.26$ & TH & He-sdOB\\
LAMOST J054711.59+170919.5 & $7.3$ & $213.9$ & $12.9$ & $2074.6$ & $0.07$ & $9.67$ & $8.37$ & $0.29$ & TH & He-sdOB\\
ATO J089.4285+27.7808 & $215.1$ & $196.7$ & $14.9$ & $2195.6$ & $0.61$ & $23.07$ & $5.66$ & $0.71$ & H & He-sdOB\\
LAMOST J060000.23+112836.6 & $-56.1$ & $213.6$ & $-28.0$ & $2030.4$ & $0.19$ & $10.85$ & $7.42$ & $0.81$ & TH & He-sdOB\\
EC 05593-5901 & $-39.2$ & $249.0$ & $57.7$ & $2148.3$ & $0.25$ & $12.93$ & $7.80$ & $2.60$ & TK & He-sdOB\\
Lan  11 & $-80.0$ & $248.8$ & $-6.5$ & $2211.4$ & $0.25$ & $12.72$ & $7.66$ & $0.20$ & TH & He-sdOB\\
LAMOST J062830.43+112018.4 & $4.8$ & $237.8$ & $8.4$ & $2313.4$ & $0.02$ & $10.38$ & $9.91$ & $0.13$ & TH & He-sdOB\\
LAMOST J062836.51+325031.5 & $-11.9$ & $198.0$ & $-17.1$ & $1855.7$ & $0.18$ & $9.70$ & $6.70$ & $0.44$ & TH & He-sdOB\\
LAMOST J063650.09+291925.0 & $26.0$ & $225.9$ & $13.2$ & $2745.2$ & $0.06$ & $13.23$ & $11.68$ & $0.90$ & TH & He-sdOB\\
LAMOST J064701.65+690625.2 & $58.5$ & $187.0$ & $5.6$ & $2113.6$ & $0.24$ & $12.39$ & $7.53$ & $1.76$ & TK & He-sdOB\\
LAMOST J070828.40+442534.8 & $-1.1$ & $144.2$ & $-40.8$ & $1283.4$ & $0.44$ & $9.75$ & $3.80$ & $1.52$ & TK & He-sdOB\\
GALEX J071319.6+380740 & $88.5$ & $84.4$ & $3.4$ & $960.6$ & $0.65$ & $10.98$ & $2.35$ & $0.88$ & TK & He-sdOB\\
GALEX J071401.4+693321 & $-14.8$ & $-70.7$ & $-96.1$ & $-742.3$ & $0.63$ & $10.64$ & $2.44$ & $5.73$ & H & He-sdOB\\
$\rm [CW83]\, 0711+22$ & $-41.8$ & $246.9$ & $-11.5$ & $2036.6$ & $0.10$ & $9.77$ & $7.98$ & $0.28$ & TH & He-sdOB\\
GALEX J071856.2+102637 & $-17.0$ & $227.1$ & $6.3$ & $2098.4$ & $0.09$ & $9.96$ & $8.36$ & $0.28$ & TH & He-sdOB\\
GALEX J072315.6+183457 & $26.2$ & $208.3$ & $6.7$ & $2150.4$ & $0.19$ & $11.66$ & $7.86$ & $0.82$ & TH & He-sdOB\\
GALEX J072824.7+414953 & $-31.5$ & $205.0$ & $-11.2$ & $2159.8$ & $0.17$ & $11.50$ & $8.11$ & $1.15$ & TH & He-sdOB\\
GALEX J073220.1+270408 & $77.1$ & $189.6$ & $18.6$ & $1773.4$ & $0.40$ & $12.49$ & $5.34$ & $0.91$ & TH & He-sdOB\\
GALEX J074125.1+125020 & $-23.8$ & $217.2$ & $16.9$ & $2601.7$ & $0.17$ & $14.14$ & $10.12$ & $1.78$ & TK & He-sdOB\\
LAMOST J075139.26+064604.8 & $-41.0$ & $225.9$ & $-11.2$ & $1937.3$ & $0.13$ & $9.52$ & $7.38$ & $0.37$ & TH & He-sdOB\\
SDSS J075733.11+235828.1 & $29.1$ & $166.9$ & $42.9$ & $2230.8$ & $0.21$ & $13.24$ & $8.71$ & $3.69$ & TK & He-sdOB\\
GALEX J075807.5-043203 & $-94.0$ & $293.7$ & $-16.8$ & $2558.3$ & $0.45$ & $20.06$ & $7.65$ & $0.51$ & TK & He-sdOB\\
KUV 07564+3723 & $-16.1$ & $226.7$ & $-3.6$ & $2163.1$ & $0.06$ & $10.12$ & $9.05$ & $1.00$ & TH & He-sdOB\\
$\rm [CW83]\, 0825+15$ & $-55.4$ & $283.5$ & $-17.9$ & $2178.0$ & $0.12$ & $10.87$ & $8.46$ & $0.61$ & TH & He-sdOB\\
Ton  927 & $-6.5$ & $163.2$ & $3.5$ & $1611.4$ & $0.29$ & $9.60$ & $5.23$ & $0.52$ & TH & He-sdOB\\
Ton  380 & $-23.7$ & $223.8$ & $-0.9$ & $2056.3$ & $0.11$ & $10.13$ & $8.11$ & $0.95$ & TH & He-sdOB\\
PG 0902+057 & $-48.6$ & $241.3$ & $4.6$ & $1732.0$ & $0.21$ & $9.30$ & $6.12$ & $0.63$ & TH & He-sdOB\\
GALEX J094044.0+004759 & $-122.9$ & $259.4$ & $-22.6$ & $2118.2$ & $0.07$ & $10.42$ & $9.08$ & $2.25$ & TK & He-sdOB\\
PG 0950+158 & $-105.3$ & $138.6$ & $-67.6$ & $1187.3$ & $0.52$ & $11.07$ & $3.45$ & $3.10$ & TK & He-sdOB\\
Ton  554 & $-52.9$ & $-3.7$ & $-108.7$ & $695.3$ & $0.85$ & $19.28$ & $1.57$ & $4.23$ & H & He-sdOB\\
PG 1127+019 & $-85.3$ & $245.1$ & $-26.9$ & $1730.7$ & $0.11$ & $8.52$ & $6.89$ & $1.37$ & TK & He-sdOB\\
Feige  46 & $-208.3$ & $219.7$ & $77.2$ & $470.3$ & $0.80$ & $9.58$ & $1.07$ & $0.56$ & H & He-sdOB\\
Feige  49 & $-171.6$ & $208.7$ & $58.6$ & $870.6$ & $0.58$ & $8.62$ & $2.26$ & $0.71$ & H & He-sdOB\\
GALEX J120521.5+224702 & $-193.3$ & $264.9$ & $161.0$ & $1157.7$ & $0.64$ & $16.48$ & $3.66$ & $8.77$ & H & He-sdOB\\
CBS 461 & $-180.5$ & $262.7$ & $83.2$ & $437.7$ & $0.85$ & $12.71$ & $1.02$ & $4.47$ & H & He-sdOB\\
CBS  58 & $-95.0$ & $192.3$ & $180.8$ & $1032.9$ & $0.32$ & $9.70$ & $5.01$ & $6.60$ & H & He-sdOB\\
GALEX J124552.8+175111 & $-224.3$ & $188.3$ & $140.5$ & $410.8$ & $0.78$ & $9.11$ & $1.11$ & $3.85$ & H & He-sdOB\\
PG 1258+012 & $-117.2$ & $310.0$ & $76.1$ & $1762.7$ & $0.28$ & $11.08$ & $6.30$ & $2.55$ & TK & He-sdOB\\
PB   146 & $-64.2$ & $139.9$ & $-17.7$ & $1309.1$ & $0.45$ & $10.28$ & $3.93$ & $2.13$ & TK & He-sdOB\\
Cl* NGC 5139   LEID   49021 & $-179.6$ & $27.8$ & $35.1$ & $91.5$ & $0.93$ & $9.15$ & $0.32$ & $0.87$ & H & He-sdOB\\
2QZ J132542.4-014655 & $-90.3$ & $223.9$ & $10.5$ & $1406.0$ & $0.15$ & $7.81$ & $5.72$ & $2.46$ & TK & He-sdOB\\
Cl* NGC 5139   LEID   40026 & $-174.1$ & $42.6$ & $60.8$ & $262.9$ & $0.90$ & $10.46$ & $0.58$ & $1.57$ & H & He-sdOB\\
Cl* NGC 5139   LEID   31031 & $-86.5$ & $-227.3$ & $-181.5$ & $-819.1$ & $0.13$ & $9.97$ & $7.65$ & $8.87$ & H & He-sdOB\\
NGC  5139  5596 & $-156.4$ & $68.9$ & $53.3$ & $457.9$ & $0.81$ & $9.75$ & $1.04$ & $1.23$ & H & He-sdOB\\
Cl* NGC 5139   LEID   42032 & $-143.2$ & $86.8$ & $87.7$ & $762.1$ & $0.71$ & $11.76$ & $1.97$ & $3.83$ & H & He-sdOB\\
2MASS J13260202-4737234 & $-102.4$ & $-203.4$ & $-152.5$ & $-518.2$ & $0.14$ & $8.35$ & $6.29$ & $7.81$ & H & He-sdOB\\
NGC  5139  4465 & $-175.6$ & $36.7$ & $52.3$ & $197.4$ & $0.92$ & $9.95$ & $0.44$ & $1.21$ & H & He-sdOB\\
NGC  5139  6139 & $-194.9$ & $-61.8$ & $-41.7$ & $-385.2$ & $0.73$ & $7.21$ & $1.11$ & $2.75$ & H & He-sdOB\\
Cl* NGC 5139   LEID   56042 & $-100.3$ & $121.3$ & $119.9$ & $1447.2$ & $0.52$ & $15.37$ & $4.89$ & $7.57$ & H & He-sdOB\\
Cl* NGC 5139   LEID   54050 & $-119.1$ & $109.0$ & $115.5$ & $1177.5$ & $0.58$ & $14.10$ & $3.75$ & $6.85$ & H & He-sdOB\\
\hline
\end{tabular}
\end{table*}

\begin{table*}

\setlength{\tabcolsep}{5pt}
\tabulinesep=1.5mm
\begin{tabular}{ L{4cm} rrrrr rrrcc}
\hline
Star & $U$ & $V$ & $W$ & $J_z$ & $e$ & R$_{a}$ & R$_p$ & Z$_{\rm max}$ &pop  &class \\
  & \kmsec & \kmsec & \kmsec & kpc \kmsec & & kpc & kpc & kpc &  &\\
\hline

Cl* NGC 5139   LEID   48062 & $-180.7$ & $41.1$ & $62.9$ & $195.7$ & $0.92$ & $9.93$ & $0.42$ & $1.46$ & H & He-sdOB\\
Cl* NGC 5139   LEID   58063 & $-128.0$ & $94.1$ & $83.1$ & $921.4$ & $0.65$ & $12.15$ & $2.54$ & $3.70$ & H & He-sdOB\\
Cl* NGC 5139   LEID   49227 & $-131.6$ & $99.4$ & $81.2$ & $862.3$ & $0.66$ & $11.21$ & $2.26$ & $3.04$ & H & He-sdOB\\
Cl* NGC 5139   LEID   56098 & $-184.5$ & $30.8$ & $56.7$ & $113.2$ & $0.95$ & $9.69$ & $0.24$ & $1.24$ & H & He-sdOB\\
NGC  5139  4547 & $-117.5$ & $112.5$ & $81.3$ & $1034.6$ & $0.60$ & $11.42$ & $2.87$ & $3.21$ & TK & He-sdOB\\
Cl* NGC 5139   LEID   34202 & $-161.3$ & $66.1$ & $55.4$ & $428.5$ & $0.82$ & $9.90$ & $0.99$ & $1.27$ & H & He-sdOB\\
Cl* NGC 5139   LEID   56132 & $-131.6$ & $92.2$ & $92.8$ & $958.5$ & $0.67$ & $13.13$ & $2.63$ & $4.82$ & H & He-sdOB\\
Cl* NGC 5139   LEID   56141 & $-105.5$ & $104.7$ & $95.4$ & $1438.7$ & $0.59$ & $16.42$ & $4.27$ & $6.41$ & TK & He-sdOB\\
Cl* NGC 5139   LEID   50311 & $-156.4$ & $72.2$ & $66.5$ & $526.1$ & $0.79$ & $10.47$ & $1.23$ & $1.89$ & H & He-sdOB\\
Cl* NGC 5139   LEID   37330 & $-199.1$ & $-12.2$ & $16.4$ & $-213.4$ & $0.88$ & $8.23$ & $0.54$ & $0.72$ & H & He-sdOB\\
NGC  5139  3595 & $-186.5$ & $-17.4$ & $-17.2$ & $-202.1$ & $0.89$ & $8.07$ & $0.46$ & $1.00$ & H & He-sdOB\\
NGC  5139  4373 & $-190.3$ & $3.3$ & $11.8$ & $-97.6$ & $0.92$ & $8.23$ & $0.34$ & $0.80$ & H & He-sdOB\\
SDSS J133449.25+041014.8 & $-185.8$ & $132.8$ & $-2.9$ & $479.7$ & $0.68$ & $7.95$ & $1.49$ & $3.76$ & H & He-sdOB\\
GALEX J134352.1+394008 & $-213.6$ & $174.0$ & $130.1$ & $139.5$ & $0.89$ & $9.25$ & $0.51$ & $6.75$ & H & He-sdOB\\
UM 610 & $-127.5$ & $139.2$ & $-35.1$ & $757.4$ & $0.55$ & $8.11$ & $2.37$ & $3.18$ & TK & He-sdOB\\
GALEX J134621.2+224835 & $-141.3$ & $249.9$ & $102.2$ & $971.1$ & $0.47$ & $9.21$ & $3.32$ & $4.16$ & H & He-sdOB\\
GALEX J135150.6+035718 & $-248.1$ & $91.4$ & $31.3$ & $243.1$ & $0.75$ & $9.26$ & $1.32$ & $7.91$ & H & He-sdOB\\
PB  1353 & $-110.2$ & $170.9$ & $61.6$ & $1080.5$ & $0.37$ & $9.53$ & $4.43$ & $5.42$ & H & He-sdOB\\
PG 1413+114 & $-98.9$ & $236.0$ & $50.8$ & $1182.0$ & $0.25$ & $7.65$ & $4.54$ & $2.85$ & TK & He-sdOB\\
GALEX J143729.2-021507 & $-59.4$ & $271.8$ & $41.5$ & $1523.9$ & $0.22$ & $8.53$ & $5.45$ & $1.43$ & TK & He-sdOB\\
GALEX J145522.9+045802 & $-62.0$ & $209.6$ & $1.7$ & $1249.4$ & $0.22$ & $7.25$ & $4.62$ & $1.84$ & TK & He-sdOB\\
GALEX J145614.5+165740 & $-274.3$ & $147.9$ & $104.2$ & $-63.9$ & $0.96$ & $7.47$ & $0.17$ & $3.75$ & H & He-sdOB\\
GALEX J152208.0+342503 & $-124.7$ & $172.4$ & $12.2$ & $1210.7$ & $0.41$ & $10.77$ & $4.53$ & $5.46$ & H & He-sdOB\\
CSO 1099 & $-53.9$ & $244.8$ & $73.3$ & $1343.7$ & $0.24$ & $8.16$ & $4.95$ & $2.36$ & TK & He-sdOB\\
SDSS J153419.42+372557.2 & $-71.7$ & $229.1$ & $64.4$ & $1373.9$ & $0.22$ & $7.97$ & $5.09$ & $2.03$ & TK & He-sdOB\\
GALEX J155202.7+261542 & $-51.7$ & $189.7$ & $0.2$ & $1035.3$ & $0.37$ & $7.43$ & $3.44$ & $2.02$ & TK & He-sdOB\\
PG 1559+048 & $-22.5$ & $242.4$ & $28.2$ & $1602.8$ & $0.16$ & $8.13$ & $5.84$ & $0.54$ & TH & He-sdOB\\
PG 1614+270 & $-107.4$ & $231.1$ & $41.7$ & $1609.4$ & $0.30$ & $10.32$ & $5.59$ & $2.55$ & TK & He-sdOB\\
GALEX J162051.0+375059 & $-42.5$ & $233.6$ & $47.7$ & $1452.2$ & $0.13$ & $7.79$ & $6.05$ & $2.41$ & TK & He-sdOB\\
GALEX J163150.9+480431 & $82.8$ & $107.3$ & $-100.3$ & $718.1$ & $0.45$ & $9.15$ & $3.44$ & $6.69$ & H & He-sdOB\\
SDSS J164042.90+311734.6 & $-207.6$ & $239.6$ & $220.8$ & $524.0$ & $0.38$ & $7.92$ & $3.54$ & $6.76$ & H & He-sdOB\\
Ton  261 & $66.4$ & $203.0$ & $-50.6$ & $1540.7$ & $0.14$ & $8.05$ & $6.11$ & $1.90$ & TK & He-sdOB\\
PG 1652+159 & $-21.5$ & $215.9$ & $21.1$ & $1284.1$ & $0.22$ & $7.13$ & $4.55$ & $1.11$ & TH & He-sdOB\\
GALEX J170045.1+391830 & $-131.8$ & $308.8$ & $92.1$ & $1578.8$ & $0.21$ & $11.79$ & $7.71$ & $7.19$ & H & He-sdOB\\
PG 1715+273 & $-27.8$ & $228.9$ & $16.0$ & $1465.2$ & $0.10$ & $7.23$ & $5.90$ & $1.46$ & TK & He-sdOB\\
UCAC4 530-064631 & $-42.1$ & $257.1$ & $24.9$ & $1685.3$ & $0.12$ & $8.42$ & $6.57$ & $1.28$ & TK & He-sdOB\\
GALEX J172034.1+145940 & $-246.4$ & $11.3$ & $-33.1$ & $-557.2$ & $0.72$ & $8.61$ & $1.43$ & $1.78$ & H & He-sdOB\\
GALEX J172533.3+523321 & $-45.1$ & $187.1$ & $-3.9$ & $1366.9$ & $0.30$ & $8.47$ & $4.54$ & $1.39$ & TK & He-sdOB\\
GALEX J181102.8+173759 & $255.4$ & $-21.6$ & $-60.2$ & $-219.9$ & $0.91$ & $13.69$ & $0.68$ & $5.02$ & H & He-sdOB\\
HS 1837+5913 & $-46.8$ & $277.4$ & $-7.2$ & $2287.4$ & $0.26$ & $13.54$ & $7.93$ & $0.84$ & TH & He-sdOB\\
2MASS J19271504+3827186 & $-27.9$ & $236.6$ & $-28.7$ & $1811.7$ & $0.10$ & $8.68$ & $7.10$ & $0.71$ & TH & He-sdOB\\
2MASS J19380163+4649446 & $0.0$ & $331.0$ & $56.6$ & $4101.1$ & $0.37$ & $31.64$ & $14.59$ & $7.55$ & TK & He-sdOB\\
2MASS J19565555+4350171 & $-8.7$ & $271.2$ & $-17.2$ & $2195.7$ & $0.17$ & $11.51$ & $8.15$ & $0.35$ & TH & He-sdOB\\
BPS CS 22940--0009 & $1.2$ & $138.3$ & $-20.9$ & $873.5$ & $0.43$ & $6.79$ & $2.70$ & $1.38$ & TK & He-sdOB\\
GALEX J211000.6+152913 & $-13.3$ & $165.6$ & $17.9$ & $1194.3$ & $0.35$ & $7.71$ & $3.75$ & $0.84$ & TH & He-sdOB\\
GALEX J211148.7+132529 & $66.1$ & $-96.7$ & $-145.9$ & $-526.0$ & $0.56$ & $8.35$ & $2.35$ & $5.95$ & H & He-sdOB\\
SDSS J212330.46+004238.8 & $33.4$ & $196.4$ & $50.5$ & $1359.7$ & $0.18$ & $7.54$ & $5.19$ & $1.90$ & TK & He-sdOB\\
GALEX J213054.5-004116 & $-18.6$ & $210.9$ & $-0.3$ & $1409.2$ & $0.13$ & $7.35$ & $5.69$ & $1.92$ & TK & He-sdOB\\
PG 2129+151 & $62.9$ & $224.1$ & $-70.2$ & $1803.0$ & $0.17$ & $9.70$ & $6.83$ & $1.75$ & TK & He-sdOB\\
GALEX J215541.6+300340 & $-15.2$ & $192.8$ & $21.3$ & $1559.0$ & $0.21$ & $8.40$ & $5.47$ & $0.71$ & TH & He-sdOB\\
GALEX J215956.5+141804 & $-42.2$ & $260.4$ & $23.2$ & $2014.9$ & $0.17$ & $10.61$ & $7.54$ & $1.11$ & TH & He-sdOB\\
PG 2204+071 & $8.3$ & $238.0$ & $-26.5$ & $1858.5$ & $0.03$ & $8.38$ & $7.90$ & $1.03$ & TH & He-sdOB\\
PG 2218+052 & $40.2$ & $201.0$ & $-52.5$ & $1601.4$ & $0.18$ & $8.51$ & $5.93$ & $1.34$ & TK & He-sdOB\\
FBS 2255+404 & $-44.0$ & $193.7$ & $15.5$ & $1797.8$ & $0.24$ & $10.10$ & $6.24$ & $0.85$ & TH & He-sdOB\\
PG 2321+214 & $70.9$ & $233.7$ & $-67.9$ & $1901.5$ & $0.25$ & $11.29$ & $6.78$ & $2.08$ & TK & He-sdOB\\
LAMOST J232812.94+295333.5 & $-127.0$ & $-26.9$ & $55.3$ & $-329.2$ & $0.86$ & $12.19$ & $0.95$ & $4.27$ & H & He-sdOB\\
GALEX J232856.6+492806 & $-25.0$ & $194.3$ & $-32.4$ & $2138.2$ & $0.16$ & $11.28$ & $8.17$ & $1.41$ & TK & He-sdOB\\
Ton S 103 & $65.5$ & $193.9$ & $164.4$ & $1475.3$ & $0.18$ & $10.67$ & $7.45$ & $6.56$ & H & He-sdOB\\
\hline
\end{tabular}
\end{table*}

\label{lastpage}
\end{document}